\documentclass[sigconf,9pt]{acmart}  

\makeatletter
\def\@ACM@checkaffil{
    \if@ACM@instpresent\else
    \ClassWarningNoLine{\@classname}{No institution present for an affiliation}%
    \fi
    \if@ACM@citypresent\else
    \ClassWarningNoLine{\@classname}{No city present for an affiliation}%
    \fi
    \if@ACM@countrypresent\else
        \ClassWarningNoLine{\@classname}{No country present for an affiliation}%
    \fi
}
\makeatother

\renewcommand\footnotetextcopyrightpermission[1]{} 
\usepackage{url}
\usepackage{graphicx}
\usepackage{caption}
\usepackage{subcaption}
\usepackage[font=small,skip=0pt]{caption}
\usepackage{setspace}
\usepackage{color}
\usepackage{soul}
\usepackage{listings}
\usepackage{multirow}
\usepackage{rotating}
\usepackage{subcaption}
\usepackage{hyperref}
\usepackage{hyperref}
\usepackage{url}

\usepackage{todonotes}
\usepackage{comment}
\usepackage{wrapfig}
\usepackage{microtype}
\usepackage{textcomp}
\usepackage{amsmath}
\usepackage{xspace}

\newcommand{\ifrmodel}{model 1}
\newcommand{\ifrmodelscaled}{model 1+}
\newcommand{\adsmodel}{model 2}
\newcommand{\pmem}{BYA-SCM}
\newcommand{\optane}{BLA-SCM}
\newcommand{\flash}{Nand Flash SSD}
\newcommand{\basecfg}{Baseline}
\newcommand{\sysOne}{configNand}

\newcommand{\sysTwo}{configBLA}
\newcommand{\sysThree}{configBYA-1}
\newcommand{\sysFour}{configBYA-2}
\newcommand{\sysFive}{configSCM}
\newcommand{\ourway}{MTrainS}
\newcommand{\etal}{et al.}

\begin{document}

\title{{\ourway: Improving DLRM training efficiency using heterogeneous memories}}
\vspace{10mm}

\vspace{10mm}
\author{Hiwot Tadese Kassa$^1$  Paul Johnson$^1$  Jason Akers$^1$ Mrinmoy Ghosh$^1$  Andrew Tulloch$^1$    Dheevatsa Mudigere$^1$  Jongsoo Park$^1$  Xing Liu$^1$   Ronald Dreslinski$^2$  Ehsan K. Ardestani$^1$ }
\affiliation{ }
\affiliation{$^1$Meta Platforms, Inc.  $^2$University of Michigan }
\affiliation{ }

\begin{abstract}

Recommendation models are very large, requiring terabytes (TB) of memory during training. In pursuit of better quality, the model size and complexity grow over time, which requires additional training data to avoid overfitting. This model growth demands a large number of resources in data centers. Hence, training efficiency is becoming considerably more important to keep the data center power demand manageable. In Deep Learning Recommendation Models (DLRM), sparse features capturing categorical inputs through embedding tables are the major contributors to model size and require high memory bandwidth. In this paper, we study the bandwidth requirement and locality of embedding tables in real-world deployed models. We observe that the bandwidth requirement is not uniform across different tables and that embedding tables show high temporal locality. We then design \ourway, which leverages heterogeneous memory, including byte and block addressable Storage Class Memory for DLRM hierarchically. \ourway~ allows for higher memory capacity per node and increases training efficiency by lowering the need to scale out to multiple hosts in memory capacity bound use cases. By optimizing the platform memory hierarchy, we reduce the number of nodes for training by 4-8$\times$, saving power and cost of training while meeting our target training performance.
\end{abstract}

\maketitle
\pagestyle{plain}

\vspace{-2mm}
\section{Introduction}


Recommendation models are broadly deployed in big technology companies to personalize the experience of their audience. For example, Google uses such models for personalized advertisements \cite{goog_widedeep}, Amazon and Alibaba for recommending items in their catalogs \cite{amazon_reco,Alibaba}, Microsoft for recommending news to users\cite{deeprecomMR}, and Meta for ranking and click-through prediction ~\cite{dlrm}.

Recommendation models are very large, requiring Terabytes (TB) of memory during training~\cite{mudigere2021softwarehardware}, and 100s of Gigabytes (GB) during inference~\cite{ardestani2021supporting}. Accelerator-enabled platforms such as GPU-enabled systems~\cite{DGX}, with 10s to 100s of accelerators, are commonly used to train such models~\cite{zhao2020distributed, mudigere2021softwarehardware}. These models take a significant amount of resources in data centers. For example, recommendation model training consumes over 50\% of AI training resources at Meta~\cite{naumov2020deep}. In pursuit of better recommendation quality, both model size and complexity are increasing by more than 1.5$\times$ year over year~\cite{jouppi2021ten}, which requires additional training data to avoid overfitting. Training models with $n$ times more parameters, requiring $m$ times more data, at the same speed, increase the resource and power demand at $O(n\times m)$. Hence, improving the efficiency of recommendation model training to manage the resource and power demand is becoming increasingly important in data centers.

Deep Learning Recommendation Models (DLRM) are neural network-based personalization and recommendation models~\cite{dlrm}. In DLRM, sparse features capturing categorical inputs through embedding tables are the major contributor to the TB scale model size, while dense features composed of Multi-Layer Perceptron (MLP) contribute to the model compute complexity. In addition to significant memory capacity requirements, sparse features may require high memory bandwidth.
Due to DLRM's considerable resource requirements and growth rate, the typical solution to accommodate these models in data centers is to scale out the model to multiple hosts. We can categorize reasons for scaling out a DLRM deployment beyond a single host at Meta as follows:

\begin{figure}[t!]
\centering
 \includegraphics[width=0.44\textwidth, trim={0.15cm 5.5cm 6.7cm 0.05cm },clip]{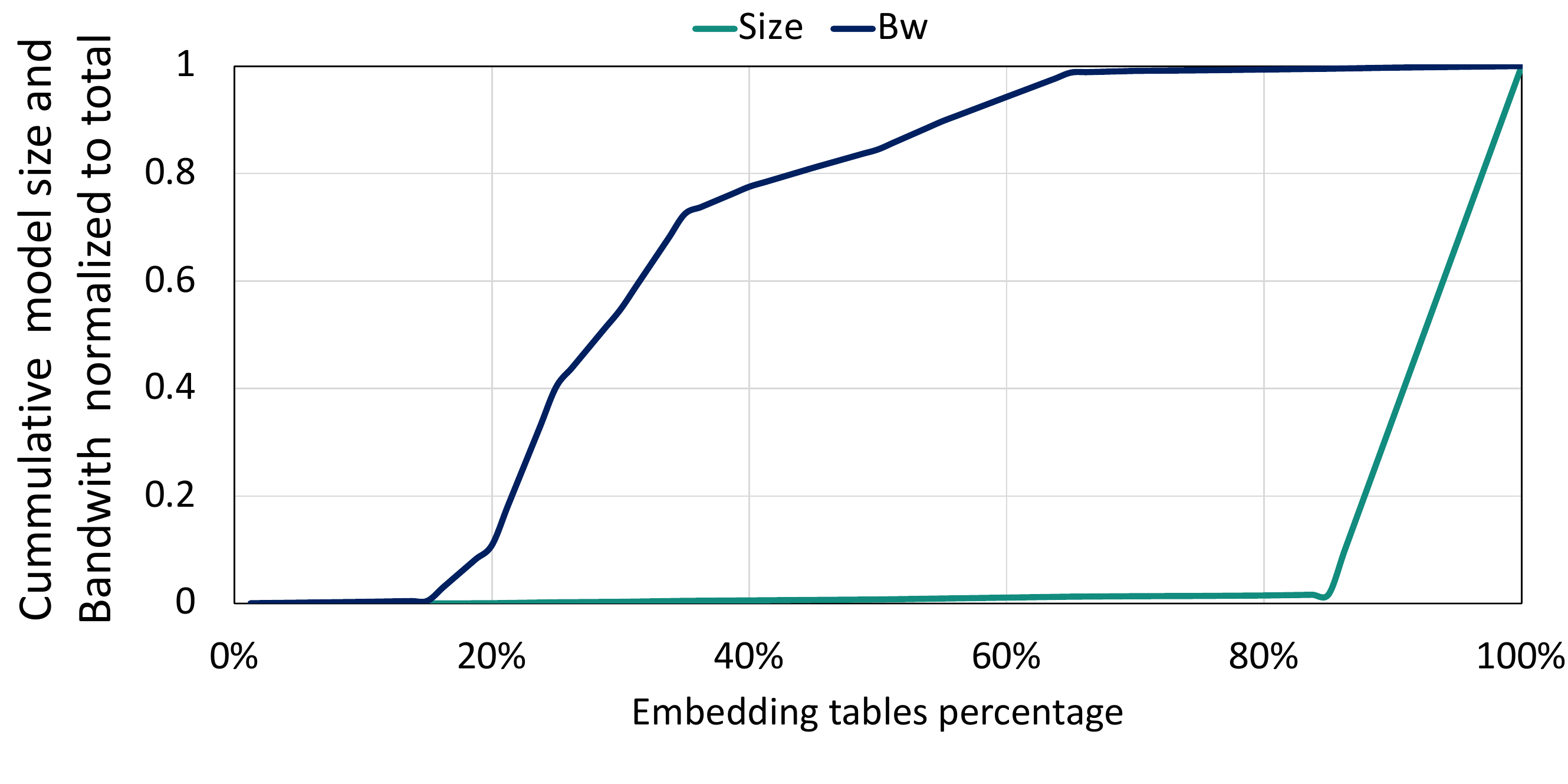}
 \caption{\footnotesize{Cumulative bandwidth and Memory size for one of the real world models we evaluate.
 }}
 \label{fig:bw_size}
\end{figure}

\begin{itemize}

  \item \textbf{Memory capacity-bound}: 
The training model size (model parameters, optimizer states, and activations) dictates the minimum number of hosts (HBM and DRAM) allocated to serve the memory size demand.
  
\item \textbf{Compute-bound/bandwidth bound}:  Given the compute intensity (e.g., in terms of Petaflops/s-day) and/or memory bandwidth requirements of the model, a number of GPUs (hosts) are used to scale the training speed (e.g., Query Per Second) to the desired target.

\end{itemize}



Diverse configurations of DLRM workloads falling into the above two categories are developed and deployed in production. Models in each category utilize the underlying hardware differently, exhibiting unique challenges to improve performance and efficiency. One such class of configurations is those that are memory capacity bound. In this case, the degree of scale-out is guided primarily by the maximum number of model parameters each host can contain. Figure~\ref{fig:bw_size} shows the cumulative memory size and bandwidth requirement normalized to the total capacity for one of the most significant capacity-bound model use case in our data center. In Figure~\ref{fig:bw_size}, each embedding table has a size calculated from the number of rows and embedding dimension. In addition, by multiplying the table's pooling factor (per-table average number of rows read per sample) by the embedding dimension size and target query per second (QPS), we can calculate the bandwidth requirement for each table. Interestingly, in the figure, the majority of the larger capacity embedding tables have a relatively low bandwidth requirement, and the tables that contribute to the most bandwidth have small sizes. By taking advantage of a hierarchy of heterogeneous memories such as high bandwidth memory (HBM), DRAM, and Storage Class Memory (SCM), it is possible to address both capacity and bandwidth requirements using fewer hosts. This increases training efficiency by limiting the need to scale out to multiple hosts only when the computation/bandwidth requirements warrant it. 

Storage Class Memories (SCMs) are technologies with the properties of both memory and storage. SCM complements HBM and DRAM by providing memory at a unique capacity, bandwidth, power, and cost target. SCM can be utilized as byte-addressable (using DIMM, or Compute Express Link (CXL)\cite{cxl} in the future), which provides $\sim$5$\times$ increase in capacity at latency and bandwidth close to that of DRAM \cite{DCPMM}, or as block addressable (using NVMe) with $\sim$20$\times$ higher capacity but at lower latency and bandwidth \cite{optane}. This flexibility allows the various memory technologies to cover a wide range of training system solutions by balancing bandwidth and capacity per host. However, such a design is not without challenges. The higher latency and lower bandwidth of the denser memory types must be accounted for when developing any scheme that distributes model parameters across different memory types. 

Previous works have shown the benefit of SCMs to complement DRAM \cite{scm_block,optane_characterstics_a,optane_early_evaluation,nvm_sim,optane_ssd_characterstic}. But little is known about the challenges and benefits of these technologies in commercial data centers for recommendation systems. Recent works \cite{eisenman2019bandana,dlrm_ssd} studied how we can use block-addressable storage for embedding tables but focused on inference, which has a different size and memory bandwidth demand than training, and as a result, imposes distinct challenges on the hardware. Previous studies also focus on using block-addressable storage with CPU. GPU-based accelerators are becoming commonplace for training recommendation systems. Hence, it is crucial to study how SCMs with lower bandwidth can be used with GPUs that have high memory bandwidth demand.

In this paper, we characterize diverse DLRM deployments at Meta and find that we have mainly capacity-bound and bandwidth-bound models. We then study these models' bandwidth, size, and locality and determine nonuniform size and bandwidth requirements across the embedding tables within a model. Our studies also show that embedding tables have low spatial locality, but high temporal locality with power-law \cite{powerlawpaper} distribution. These characteristics make our DLRM models suitable for hierarchical memory with HBM, DRAM, and byte and block addressable SCM. We then design \ourway, an end-to-end trainer that efficiently leverages a heterogeneous memory hierarchy along with GPUs. \ourway~ has a key-value \cite{rocksdb} storage system residing in SSDs for large (TB scale) embedding tables management. Then, to hide the low bandwidth and high latency of SSDs, \ourway~ implements a GPU-managed, software-based configurable hierarchical cache that uses DRAM and SCM for hot and cold embedding rows, respectively. We use two of our extensive resource-consuming models representing both capacity- and bandwidth-bound workloads for our experiments. Our results demonstrate that for capacity-bound models, by using \ourway, we can reduce the number of hosts used for training by 4$\times$\ for current models and by 8$\times$\ for future scaled models while meeting our service level agreement (SLA) QPS target. To the best of our knowledge, this is the first work that studies SCM usage in GPU-enabled systems in a commercial data center for recommendation system training. In summary, we make the following contributions: 
\begin{itemize}
    
\item Extend the memory hierarchy of DLRM training beyond HBM and DRAM to SCM in byte and block addressable forms.
   
\item Characterize and experiment on large scale production-based real workloads, and discuss the real world scenarios where such hierarchical memory can win efficiency. 
 
 \item Discuss the system-level performance trade-offs of byte and block addressable SCMs for the DLRM training usecase.
 
  \item Extend the open source DLRM with different memory support to facilitate hardware/software research on larger AI models.

\end{itemize}

\vspace{-2mm}
\section{Background} \label{background and motivation}
This section discusses the architecture of deep learning recommendation models (DLRM), FBGEMM\_GPU (an optimized open-source GPU kernel library we used in our designs), and the various memory and storage components in our systems.  
\vspace{-2mm}
\subsection{Deep learning recommendation model}

Recommendation systems are widely deployed to rank content such as news feeds, videos, and products based on user preferences and interactions.
To accurately rank user preferences, recommendation systems such as DLRM \cite{dlrm} use deep neural networks. However, DLRM is unique from DNNs because it has dense and sparse features to capture user and item attributes, leading to different characterizations as far as the underlying system is concerned.
\vspace{-1mm}
\subsubsection{DLRM architecture and components}
 Figure \ref{designoveriew} shows the internal components of DLRM, the details are described below:
\begin{figure}[t!]
\centering

\includegraphics[width=0.37\textwidth, trim={0.05cm 7.0cm 17.7cm 0.05cm },clip]{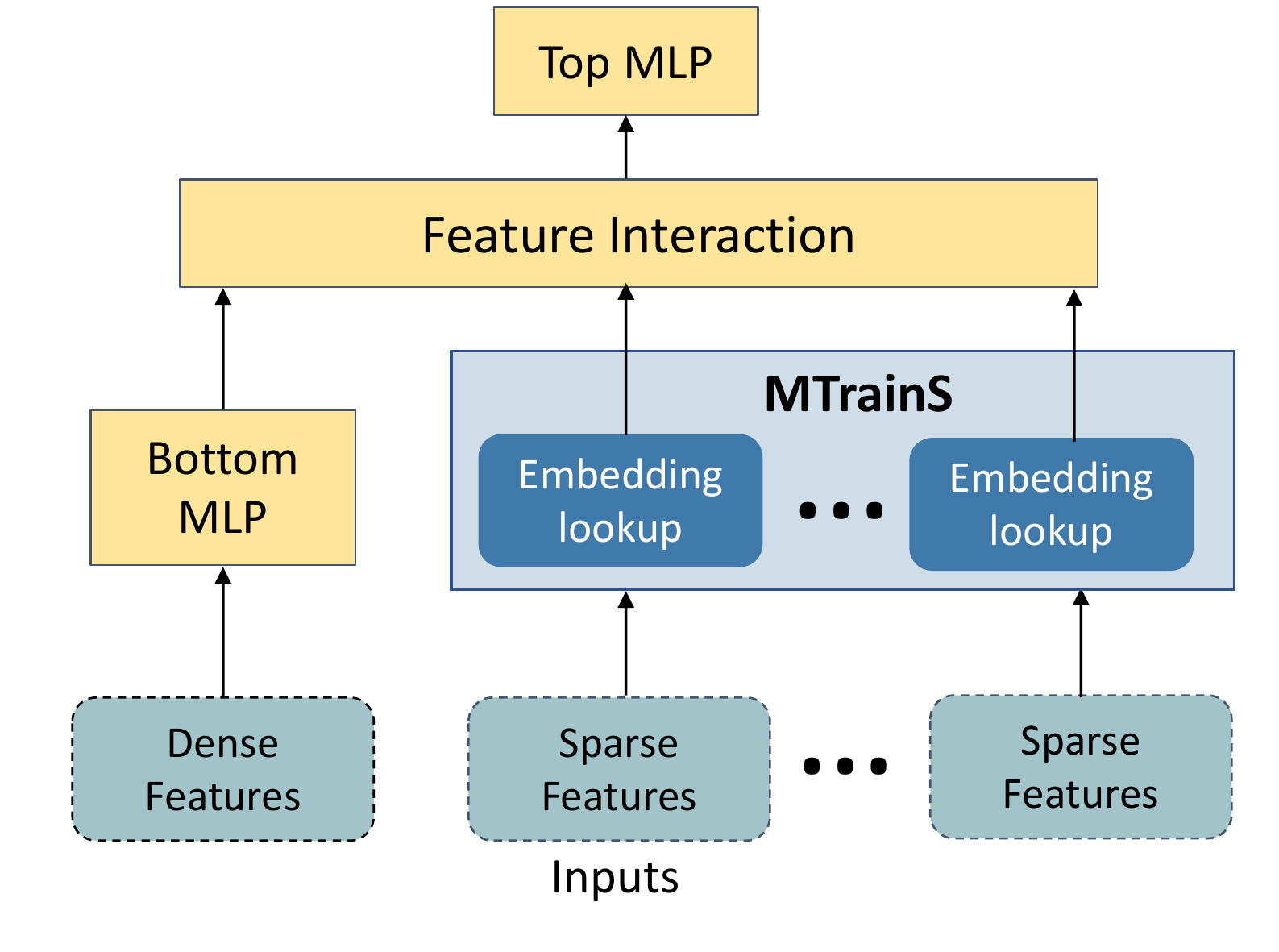}
 \caption{\footnotesize{DLRM architecture and where \ourway~fits in DLRM.}}
 \label{designoveriew}
\end{figure}
\begin{table*}
\centering
\footnotesize
\begin{tabular}{|l|c|c|c|c|c|} 
\hline
       \textbf{Characteristics}            & \textbf{\flash~} & \textbf{Optane SSD} & \textbf{DRAM} & \textbf{Optane memory} & \textbf{HBM} \\ 
\hline


Power (mW / GB) (mW / GB/ s for HBM )        & 5.7 & 35      &  375    & 98   &  5000   \\\hline

 Cost per GB  relative to \flash~
 &1        & 10.4&  68.8   & 26.5    &   -     \\\hline

Granularity of access       & block &block      &   byte      &  byte    & byte     \\\hline

Total capacity per host (GB) /Total BW per host (GB/s) &8192/6&2048/6&384/200 & 2048/84&320/12800 \\\hline
Endurance (DWPD)&0.8&100&-&-&- \\\hline \end{tabular}
\caption{\footnotesize{Characteristics of NAND SSD, Optane SSD, DRAM, Optane memory and HBM per module taken from product specifications. 
}}
\label{table:charactersitics}
\vspace{-3mm}
\end{table*}



\begin{figure*}[t!]
\centering
  \begin{subfigure}{0.64\columnwidth}
   \includegraphics[width=1.2\textwidth, trim={0.05cm 6.0cm 8.3cm 0.05cm },clip]{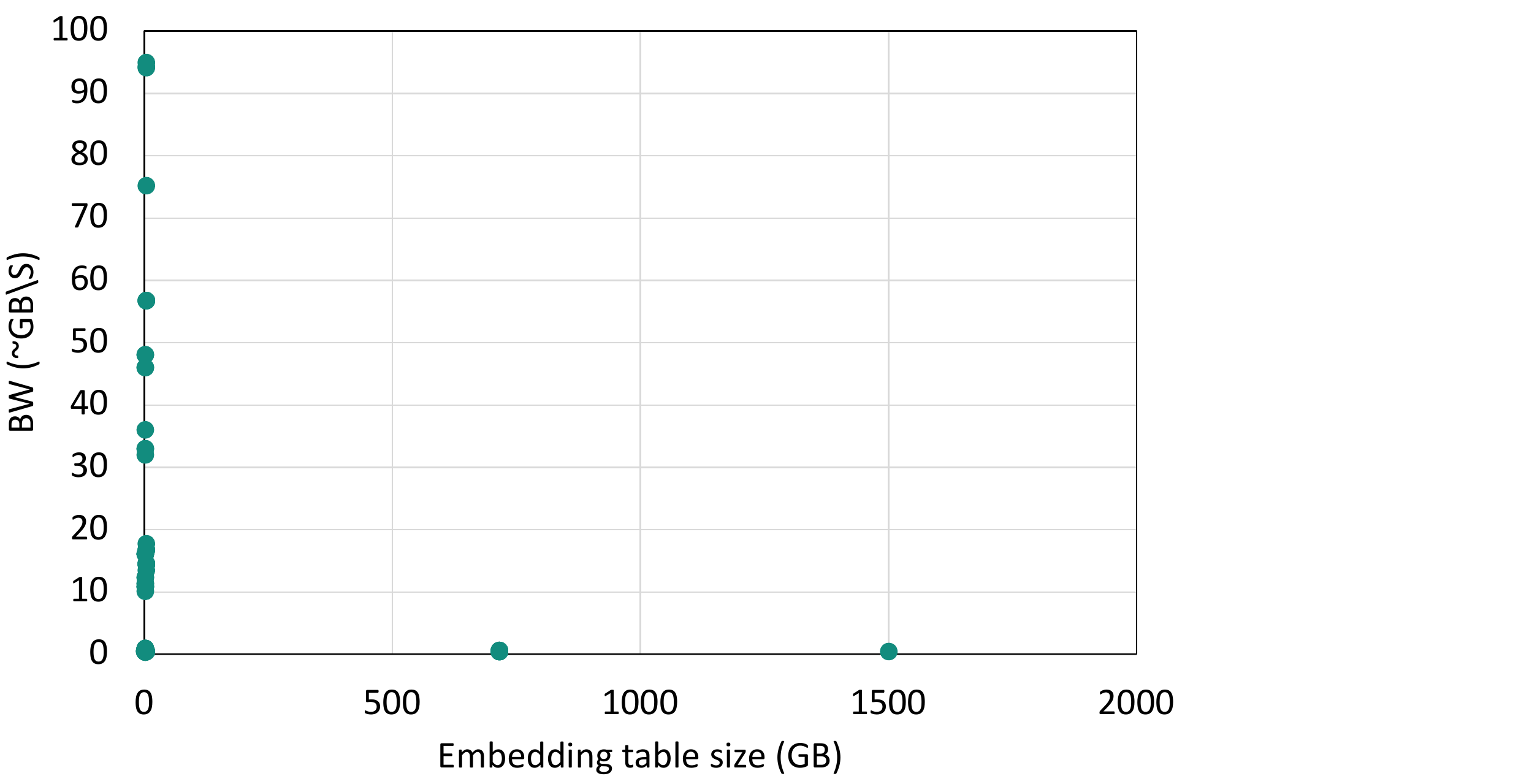}

\vspace{-2mm}
\caption{\textmd{Model 1} }
\label{figures:bw_model1}
    \end{subfigure}
\begin{subfigure}{0.64\columnwidth}
 \includegraphics[width=1.2\textwidth, trim={0.05cm 6.2cm 8.5cm 0.05cm },clip]{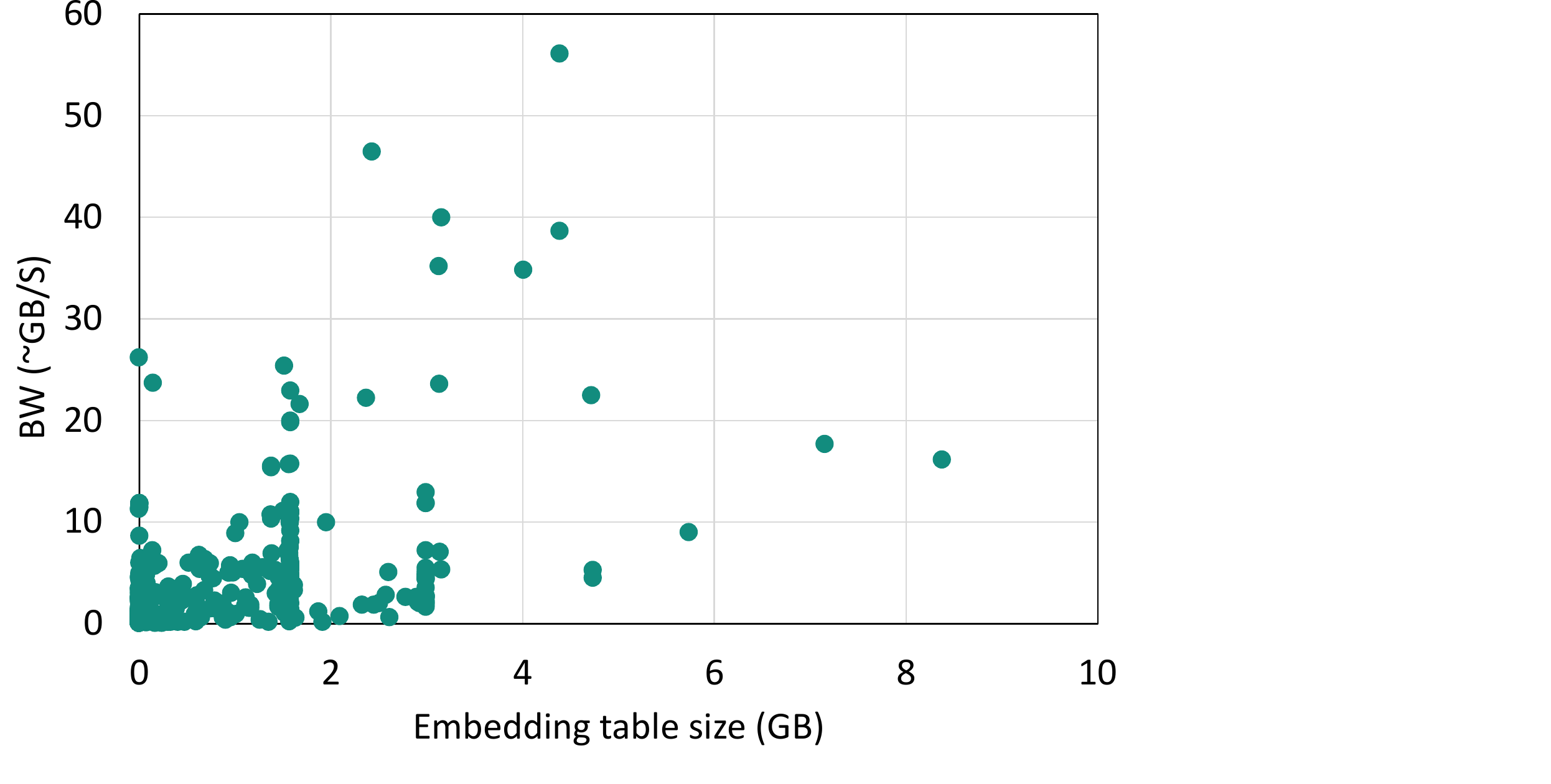}

\vspace{-2mm}
\caption{\textmd{Model 2} }
\label{figures:bw_model2}
    \end{subfigure}
   \begin{subfigure}{0.64\columnwidth}
 \includegraphics[width=1.2\textwidth, trim={0.05cm 4.8cm 8.3cm 0.05cm },clip]{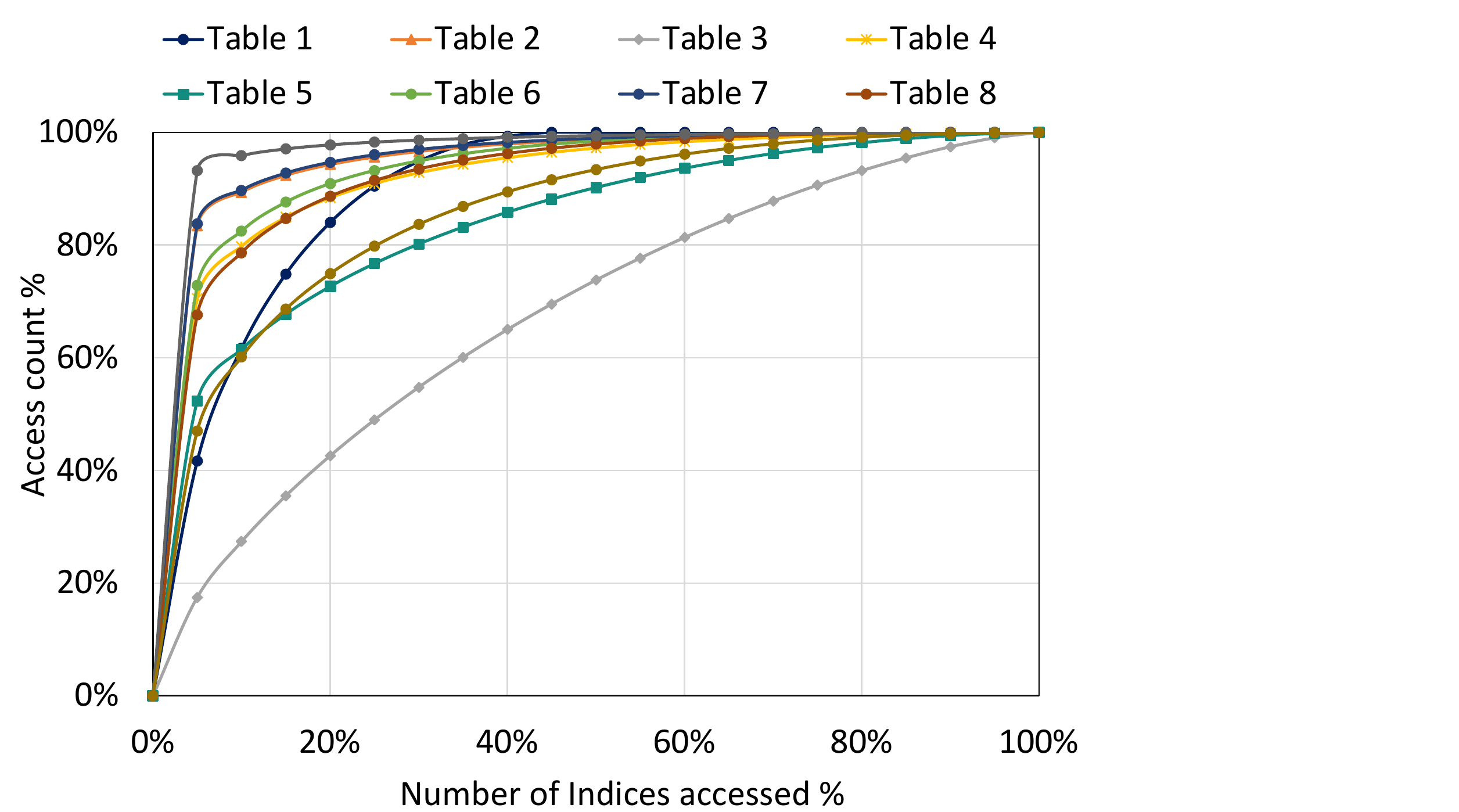}
\vspace{-5mm}
\caption{\textmd{Temporal locality analysis of embedding tables}.}
 \label{locality}
    \end{subfigure}
 \caption{\footnotesize{BW vs Size distribution of Model 1 and Model 2 and locality analysis of various embedding tables.}}
 \vspace{-4mm}
 \label{size_bw}
\end{figure*}

\noindent\textbf{Input features:} 
Users' data and products are represented by dense and sparse input features. \textbf{Sparse} features represent categorical inputs, such as a page a user likes in a list of pages. As the name suggests, this data is sparse, i.e., a user has likely interacted with a small subset of billions of pages available. \textbf{Dense} features include continuous inputs, such as user age.  

\noindent\textbf{Embedding tables:}
A naïve representation of categorical inputs would be to use binary vectors. For example, if we have 4 pages with IDs from 0 to 3 and if a user likes ID 0 and 2, the embedding vector for the user will be (1,0,1,0). But we have billions of pages, hence such representation will be very large and sparse. Furthermore, binary vectors do not represent the relationship between similar pages. To avoid these problems, DLRM uses embedding tables that map categorical features into a dense representation. In this case, categorical inputs such as a page will be represented by a short vector, and similar pages will be located closer in Euclidian space. In the embedding tables, the column represents the embedding vector, and the rows are items in a category. Within a model, there are multiple categorical features, and the number of rows varies across tables. 
Embedding table operators look up a subset of rows in an embedding table, and pool the result using \textit{sum}, \textit{mean} or \textit{max}~\cite{NEURIPS2019_9015}. 

\noindent\textbf{Bottom MLP layer:}
The continuous inputs are transformed and projected into a dense space by a bottom multi-layer preceptron (MLP) that is composed of a series of fully connected (FC) layers and activation functions.  

\noindent\textbf{Feature interaction and top MLP:}
The dense projections of categorical and continuous features are aggregated (e.g. through concatenation), and then a set of MLP layers capture the interaction between different features. 

\vspace{-2mm}
\subsubsection{Embedding tables and operators}
Embedding tables impose unique challenges in systems designed for recommendation systems. Real-world use cases of embedding tables require a large memory capacity, up to 10s of TB. Equation~\ref{eq:emb_size} formulates the memory capacity requirements for a model with $T$ embedding tables, $H$ average number of rows per table (also known as embedding table hash size), $D$ elements per row (embedding vector dimension), and element precision of $p$ bytes. 
\begin{equation}
\label{eq:emb_size}
    Memory(Sparse Parameters) = T \times H \times D \times p
\end{equation}

Given the massive size of the embedding tables, typical optimizers with a small number of states per row, such as Adagrad~\cite{lydia2019adagrad}, is commonly used for sparse features. We can rewrite Equation~\ref{eq:emb_size} to include both model parameters and optimizer states ($o$).
\vspace{-3mm}

\begin{equation}
\label{eq:emb_size_mop}
    Memory Capacity = T \times H \times (D + o) \times p
\end{equation}

Embedding tables are also memory BW intensive, as each training sample accesses multiple rows per embedding lookup. Assuming $L$ rows are accessed per table for each training sample, Equation~\ref{eq:emb_bw} formulates the BW requirement for embedding tables training to achieve a given QPS. Since both the forward and backward passes consume all the rows accessed, the equation is multiplied by 2.  
\vspace{-0.5mm}
\begin{equation}
    \label{eq:emb_bw}
    Memory BW = QPS \times T \times D \times p \times L \times 2
\end{equation}

\vspace{-2mm}

\subsection{FBGEMM\_GPU kernel library}
All software development contributions of this paper are on top of the FBGEMM\_GPU (FBGEMM GPU kernel library) \cite{fbgemm,fbgemm1}, which is a high-performance GPU CUDA operator library for deep neural network training and inference. FBGEMM\_GPU provides an efficient embedding table operator, data layout transformations, and other optimizations. It supports efficient embedding table access by providing HBM-based caching and DRAM utilization using unified virtual memory. We extended FBGEMM\_GPU to add operators for SSD/SCM-based training and for various caching mechanisms to hide the high access latency of SSD. 

%

\vspace{-2mm}
\subsection{Storage and memory types}
We examined different storage and memory technologies to accommodate larger models per node. Each unit has its benefits and drawbacks. Table \ref{table:charactersitics} shows the characteristics of each technology. The detailed descriptions are as follows:

\noindent\textbf{\flash: }It is the densest and cheapest memory technology we leverage, as seen in Table \ref{table:charactersitics}. However, these SSDs have low input/output Operations Per Second (IOPS) and significant latency compared with other memory types we evaluate. Moreover, they have limited write endurance, defined as Drive Writes Per Day (DWPD). NAND flash-based solution's limited IOPs make it best suited for a limited range of sparse features with high memory capacity and low bandwidth. 

\noindent\textbf{Optane SSD:} It is based on Intel's 3D XPoint technology. It has higher IOPS, especially for lower access granularity requests and lower latency compared to NAND flash SSD. These SSDs have balanced read and write latency as well as 100$\times$ better write endurance (DWPD) (see Table \ref{table:charactersitics}). However, these SSDs come at a 10$\times$ higher price per GB than NAND SSDs. We refer to this memory as \optane~ for BLock-Addressable SCM.

\noindent\textbf{Optane Memory (PMEM):}
Intel Optane memory sits between DRAM and SSDs in the memory hierarchy. It is a cheaper alternative memory to DRAM with a 4-8$\times$ higher density but has lower bandwidth and higher latency than DRAM. We refer to this memory as \pmem~ for BYte-Addressable SCM. \pmem~ operates in Memory Mode and App Direct Mode. In \textbf{Memory Mode} DRAM serves as a direct map cache, while \pmem~ is exposed as a single volatile memory region. The DRAM cache and \pmem~ accesses are handled exclusively by the CPU's memory controller, and applications have no control of where their memory allocations are placed (DRAM cache or \pmem~). In \textbf{App Direct Mode}, DRAM and \pmem~ are configured as two distinct memories. Here, the applications fully control read and write access to each memory. Note that the persistence characteristics of these memories are not relevant for our use case of training DLRM.

\noindent\textbf{DRAM:}
DRAM has high bandwidth, and low read/write latencies. However, as seen in Table \ref{table:charactersitics}. it is more expensive per GB, has a lower density, and has higher power consumption per GB compared to \pmem~ and SSDs. Additionally, the maximum DRAM capacity is limited by the number of DIMMs available on a host.

\noindent\textbf{HBM:}
Many modern GPUs designed for HPC/AI Training utilize a memory technology even faster than DRAM. High Bandwidth Memory (HBM) has higher memory bandwidth than conventional DRAM. These memory modules are soldered onto the GPU, so the capacity is fixed. Therefore, the size available for embedding table storage is limited per GPU.
\begin{figure*}[th!]

  \begin{subfigure}{0.7\columnwidth}

\includegraphics[width=1.0\textwidth, trim={0.05cm 6.2cm 10.5cm 0.05cm },clip]{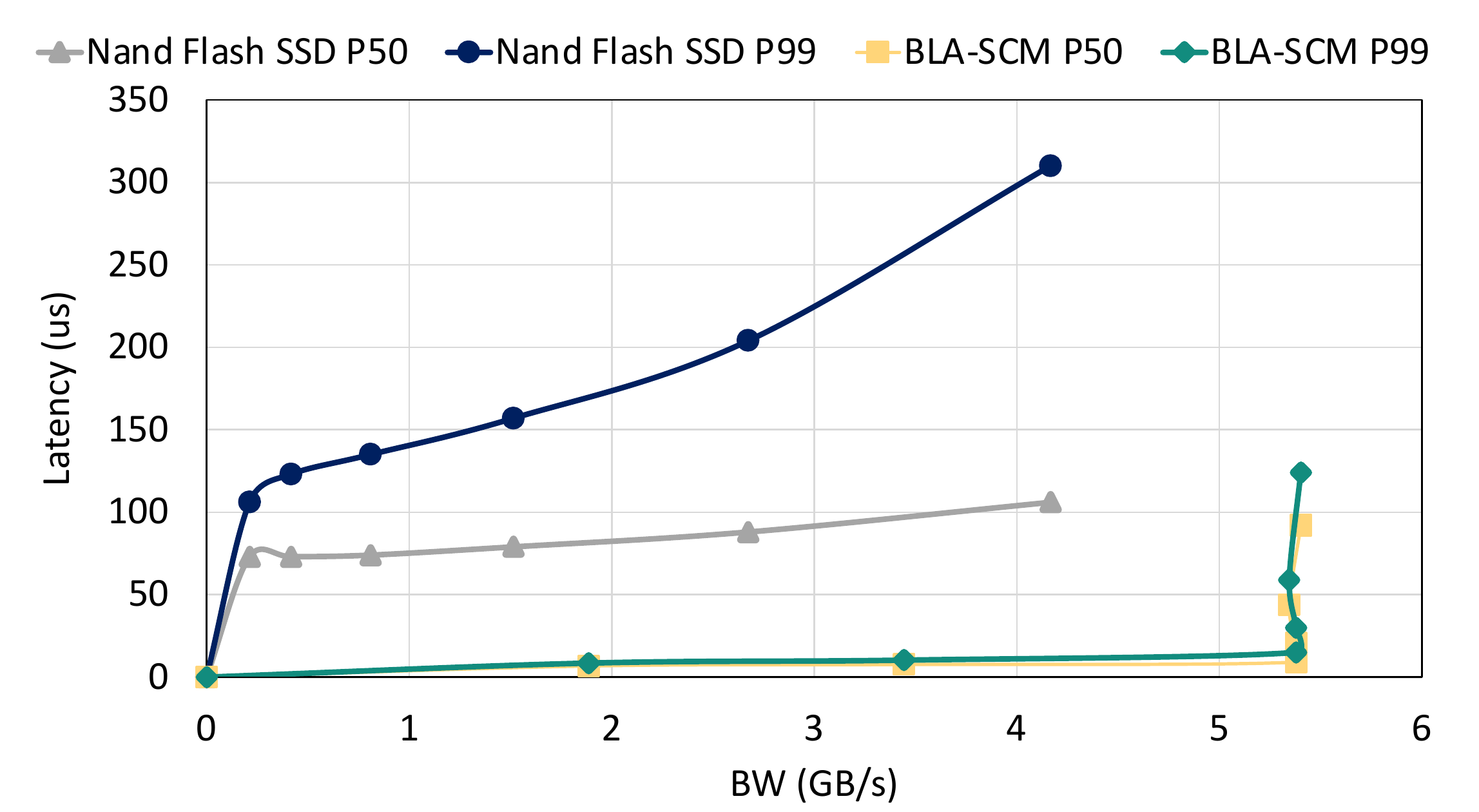}
\vspace{-5mm}

\caption{\textmd{P50, P99 latency and BW of \optane~ and NAND SSD.} }
\label{figures:storage_characterstics}
    \end{subfigure}
\begin{subfigure}{0.7\columnwidth}
 \includegraphics[width=1.0\textwidth, trim={0.05cm 6.0cm 9.3cm 0.05cm },clip]{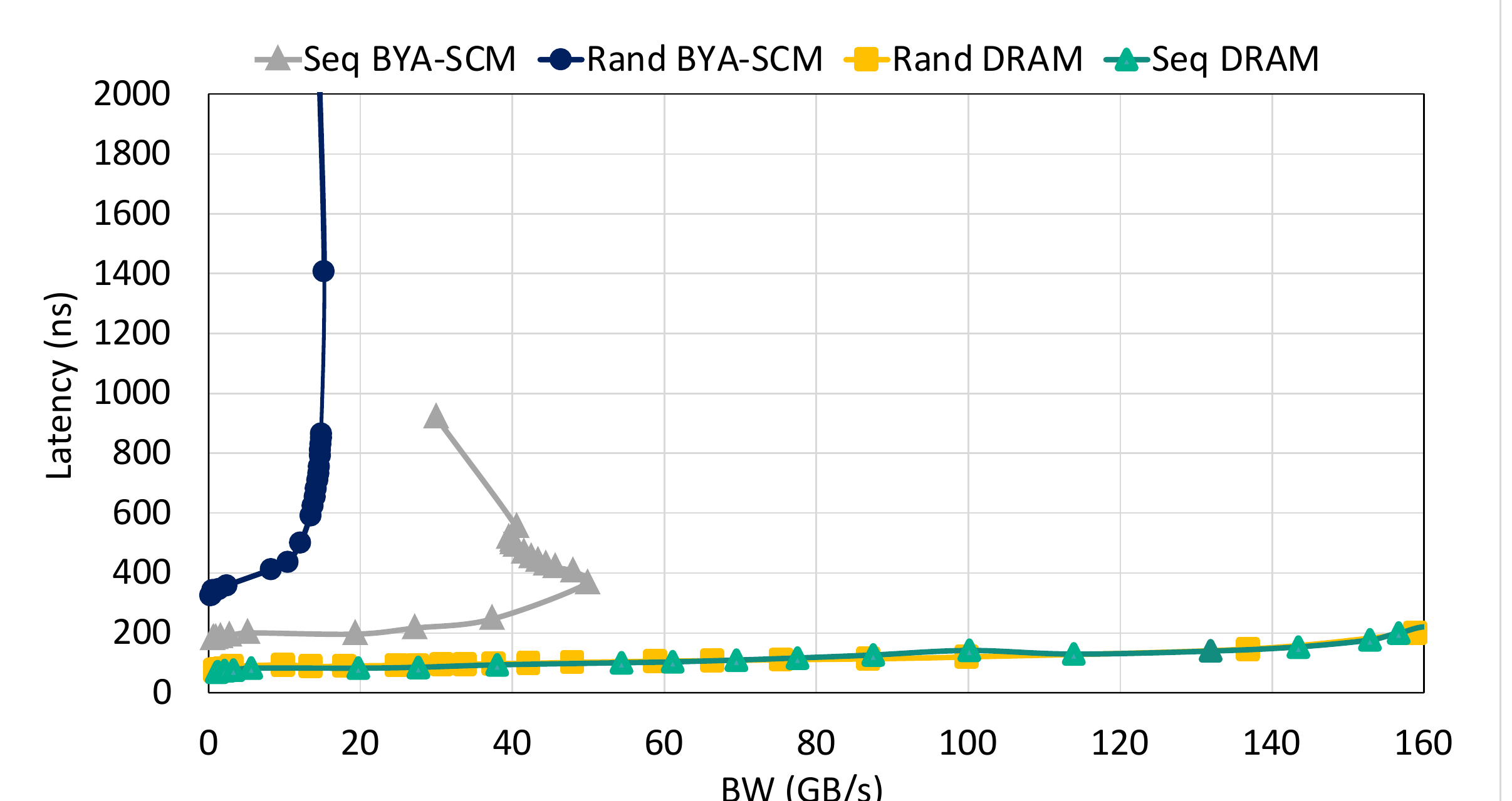}
\vspace{-5mm}

\caption{\textmd{Latency and BW of DRAM and \pmem.} }
\label{figures:memory_characterstics}
    \end{subfigure}
\begin{subfigure}{0.6\columnwidth}

\includegraphics[width=1.05\textwidth, trim={0.20cm 9.4cm 14.5cm 0.15cm },clip]{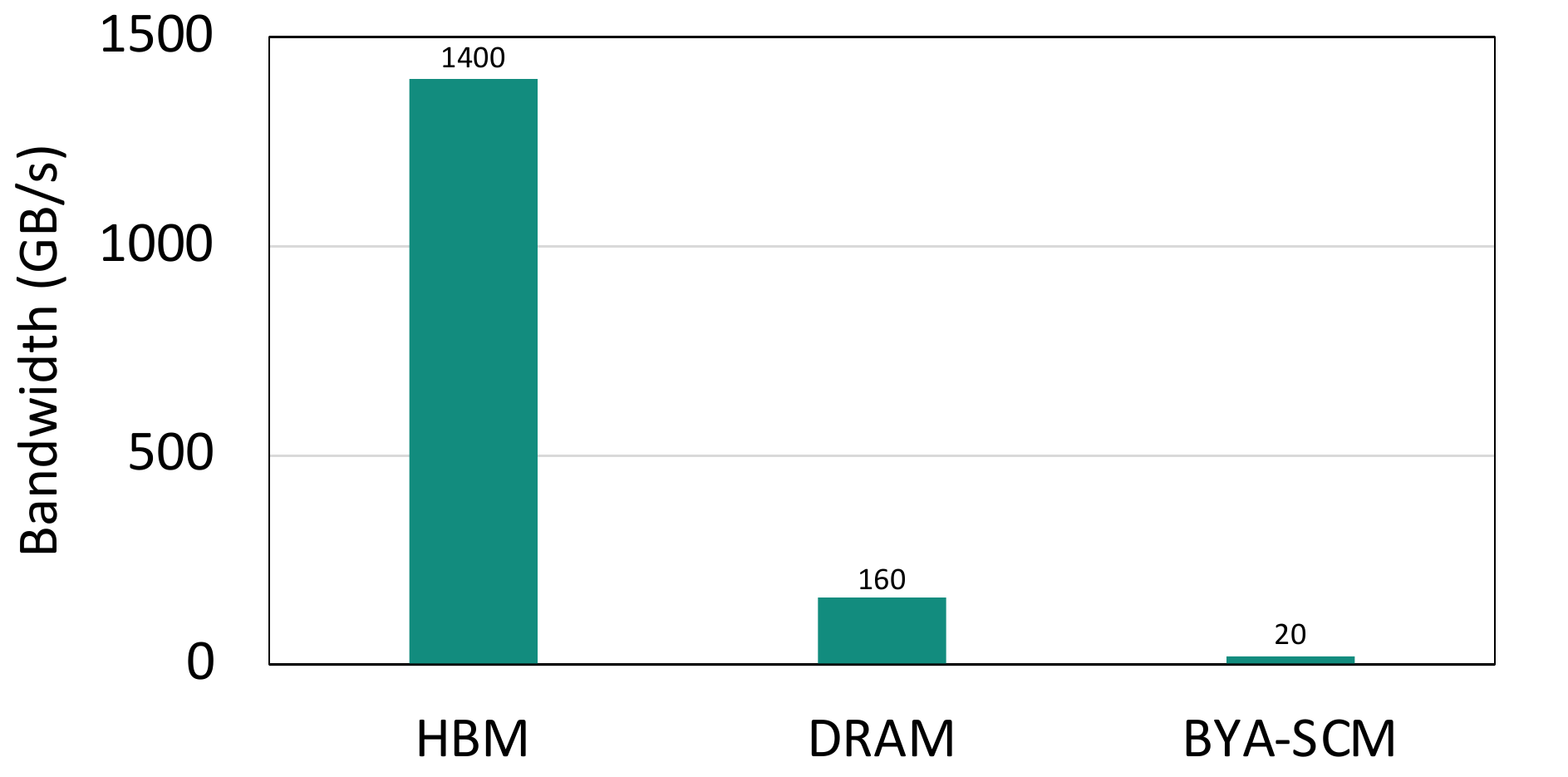}
\vspace{-5mm}

\caption{\textmd{BW comparison of HBM, DRAM and \pmem.} }
\label{figures:HBM}

    \end{subfigure}

\label{io}
\caption{
System memory and storage characterization.
}

\end{figure*}
\begin{table}[h]
\centering
\footnotesize
\begin{tabular}{|l|c|c|c|}
\hline
\textbf{  }         &  \ifrmodel & \ifrmodelscaled &    \adsmodel                  \\
\hline
Features                     &$\sim10s$ & $\sim10s$ & $\sim100s$                           \\
Total bw (GB/s) & 1300 &2600&7136\\
Embedding dimension&128&256&128  \\
Data type           & 4 byte &4 byte & 4 byte                                 \\
Average pooling factor          &33 &33 &18                            \\
Num MLP layers&7&7 &20\\      
\hline
\end{tabular}
\caption{Workload specifications.}
\label{tab:workloads}
\end{table}
\section{Workload characterization}\label{workloads_characterstics}

The large size and bandwidth requirements of DLRM workloads impose challenges in systems design. We studied the size, bandwidth, and locality of embedding tables in various deployed DLRM models in production to understand how we can improve their performance and power efficiency with heterogeneous memories. We select two of the most prominent representative models in our data center with distinct characteristics and show the details here. The models are \ifrmodel, which is used for ranking content in various services, and \adsmodel, which is used for click-through rate (CTR) prediction for user content and item recommendations. \ifrmodelscaled~ is a future scaling of \ifrmodel~ with similar BW and locality characteristics as \ifrmodel. These models have TB scale sizes. The characteristics are shown in Table  \ref{tab:workloads}.
\vspace{-1mm}
\subsection{Bandwidth and size distribution} \label{section:bw}
Figure \ref{figures:bw_model1} and \ref{figures:bw_model2} show a sample of the bandwidth and size distributions of embedding tables found in \ifrmodel~ and \adsmodel. We use Equation \ref{eq:emb_size} and \ref{eq:emb_bw} to study size and bandwidth. We use the tables' row numbers, dimensions, and precision from production configurations to calculate the size of embedding tables. We then use the acceptable QPS (SLA) for training each model in our data center and the pooling factor to calculate bandwidth. Pooling factor is determined from a large historical data of how many times each embedding table is accessed per lookup. As seen in the figures, the bandwidth vs. size distributions for the two models are distinctive. In \ifrmodel, we have smaller size embedding tables with high bandwidth requirements and large size embedding tables with lower bandwidth requirements. These types of tables fit intrinsically to a hybrid memory system with both large size/low bandwidth and small size/large bandwidth components. The cumulative bandwidth for our target QPS for \ifrmodel~ can be satisfied by a single HBM+DRAM system.
Here, we scale out the model to multiple hosts to fit the model parameters with more memory capacity. 
On the contrary, in \adsmodel, there are considerably more embedding tables that vary significantly in size and bandwidth. In this case, we scale models to multiple hosts to increase both the memory capacity and bandwidth. 
 Due to this wide variance in size/bandwidth requirements, bandwidth and size distribution of models are key factors in DLRM memory hierarchy designs.

\vspace{-2mm}
\subsection{Locality in embedding table lookups}\label{section:locality}

As discussed in \cite{ardestani2021supporting}, the spatial locality in embedding tables is very low because embedding tables represent sparse categorical features, and the embedding row access is very irregular. However, we have a considerable temporal locality that makes caching effective in our trainer design. To investigate temporal locality, we examine the frequency of access of embedding table Indices for several embedding tables of \ifrmodel~ and \adsmodel~running in production for 24 hours. We show the results for the representative embedding tables in Figure \ref{locality}. The figure shows that access to most tables follows a power-law distribution. As also shown in~\cite{ardestani2021supporting}, we observe  80\% of the indices accessed come from 10\%-40\% of the total Indices for most tables. Hence, we can take advantage of heterogenous memories and storage by placing colder embedding tables and embedding rows in large but slower memories like \pmem, \optane, or \flash, while hot embedding tables and embedding rows can still enjoy faster but smaller size memories like HBM and DRAM through caching because of the high temporal locality in DLRM workloads. In our design in Section ~\ref{sec:design}, we emphasize how we can maximize caching to hide latency and provide higher BW by using hierarchical caching.



\section{System design challenges and considerations}
\label{sec:memory_characterstics}
This section discusses the challenges of adopting heterogeneous memories and storage for DLRM training and our considerations in  the workloads and hardware characteristics for our designs. 
\vspace{-2mm}
\subsection{Memory Performance evaluation} 
We compare the latency and BW of the memories and storage to understand how they fit with our workload characteristics. 
We measure the \optane~and \flash~latency and BW with FIO~\cite{axboe2013fio} for random read workload with different queue depths to increase BW utilization. In Figure \ref{figures:storage_characterstics}, \optane~has a latency in $\sim$10{\textmu}s range for both P50 and P99 at similar BW, whereas \flash~has a latency of ~100{\textmu}s, and P99 latency is significantly higher than P50. Also, note that increasing BW utilization in \flash~ increases the access latency. This shows that for \flash~ we have to be careful with the BW utilization to prevent significant latency. Given the high temporal locality in our models, it is advantageous to implement caching to reduce SSD traffic, especially with the performance limitations of \flash.


We use Intel's Memory Latency Checker (MLC) \cite{mlc} to measure DRAM and \pmem~ latency and BW. We use sequential and random read workloads with different memory traffic rates. In Figure \ref{figures:memory_characterstics}, \pmem~ achieves $\sim$15 GB/s and DRAM 170 GB/s BW. Further, \pmem's~ latency increases with increased memory traffic ($\sim$200ns - 800ns for a sequential and $\sim$350ns - 1500ns for a random read), and the BW saturates at high traffic. Then the latency increases with no BW change. However, DRAM has a much lower latency and higher BW than \pmem, and it maintains the same latency for sequential and random access. While spatial locality in our workloads is low in the 4KB block access granularity of SSDs, for \pmem, the sequential access granularity is 256 bytes. The access granularity of an embedding lookup in our workloads is 512-1024B. Hence, we can still achieve \pmem~ sequential access performance. However, because of the BW and latency differences between DRAM and \pmem, the most practical design is a hierarchical configuration, where DRAM is used for hot embedding rows and \pmem~ for colder ones. We should also consider the traffic to \pmem~ to avoid large latency and BW saturation.   


We use gpumembench\cite{gpumembench} for HBM and MLC for DRAM and \pmem~ to compare BW differences. In Figure \ref{figures:HBM}, we see that HBM has significantly higher BW than  DRAM and \pmem. These BW and size differences (as seen in Table \ref{table:charactersitics}) show that we need to optimize the placement of embedding tables and rows to these memories to maximize BW utilization. 





\subsection{IOPS vs BW}
\label{sec:iops_vs_bw}
Although we have temporal locality in the embedding tables, adjacent rows are accessed in a non-sequential manner and lack spatial locality. Additionally, the embedding dimensions typically range from 64 to 256 (i.e., 256-1024 bytes with single precision). When considering block addressable technologies such as \optane~ and \flash, each access to an embedding row could consume less than the block size (e.g., 4KB), resulting in a waste of BW (referred to as read amplification). To account for the impact of read amplification, we track IOPS instead of BW for the block addressable units and study how we should use these memories to satisfy IOPS demand. Equation~\ref{eq:iops} formulates the required IOPS, assuming $T_B$ tables are placed on SSD, with an average pooling factor of $L_B$. $\alpha$ is used to factor in the locality of accessing embedding tables, which reduces access to the lower level block addressable memory.
\begin{equation}
    \label{eq:iops}
    IOPS = QPS \times T_B \times L_B \times \alpha  
\end{equation}

The IOPS requirement for \ifrmodel~ and \adsmodel~ to accommodate the entire model in one node while placing $T_B$ tables on SSD for our target QPS is 6.25M and 75M, respectively, without considering the locality. Typically, SSDs have IOPS in a range of 500K-1M IOPS limit. For example, if we have a cache hit rate of 70\% for \ifrmodel, the IOPS will be 1.875M. Whereas for \adsmodel~ with 70\% locality, we still require 22.5M IOPS. Hence, the locality is not only important for optimizing the high latency of SCM, but it will help us to operate within the IOPS limit of the hardware. Therefore, we focus on maximizing the locality of the models in our designs. 
\subsection{Endurance}
\flash~ and \optane~\footnote{Optane in DIMM form factor, referred to as \pmem~ in this paper, is claimed to not be bounded by endurance} have a limited number of program/erase cycles that can be performed before a memory cell wears out. This is measured as how many times the entire drive can be written to each day of its lifetime (typically 3-5 years) which is called Drive Writes Per Day (DWPD). Equation~\ref{eq:wr_per_day} formulates the amount of data written per day during training for a given QPS, where $T_B$ tables are placed on SSD, with $L_B$ average pooling factor, $D$ elements per row, $\alpha$ locality, and element precision of $p$ bytes.
\begin{equation}
    \label{eq:wr_per_day}
    write / day = 24 \times 3600 \times QPS \times T_B \times L_B \times D \times p \times \alpha 
\end{equation}

For example we write  $\sim$10TB per day to SSDs for \ifrmodel~  and  $\sim$100TB for  \adsmodel~  while placing  $T_B$ tables in SSDs. DLRM workloads are write-intensive. When using SSDs for embedding table storage, we want to limit our writes per day below the stated DWPD limit to avoid premature drive failure.
\subsection{Workload scaling}
Deep learning models, including DLRM, are scaling rapidly in complexity and size. A number of sparse features are used in the model, and hence the number of embedding tables, along with the embedding dimensions per table, are among the main contributors to the increase in the memory capacity of DLRM workloads. Any system solution needs to be able to consider such scaling during the lifetime of the system. In this paper, we design systems for current models and test whether our design holds in future scaling. 
\begin{figure}[!t]
 \centering
   \includegraphics[width=0.37\textwidth, trim={0.4cm 8.6cm 11.0cm 0.05cm },clip]{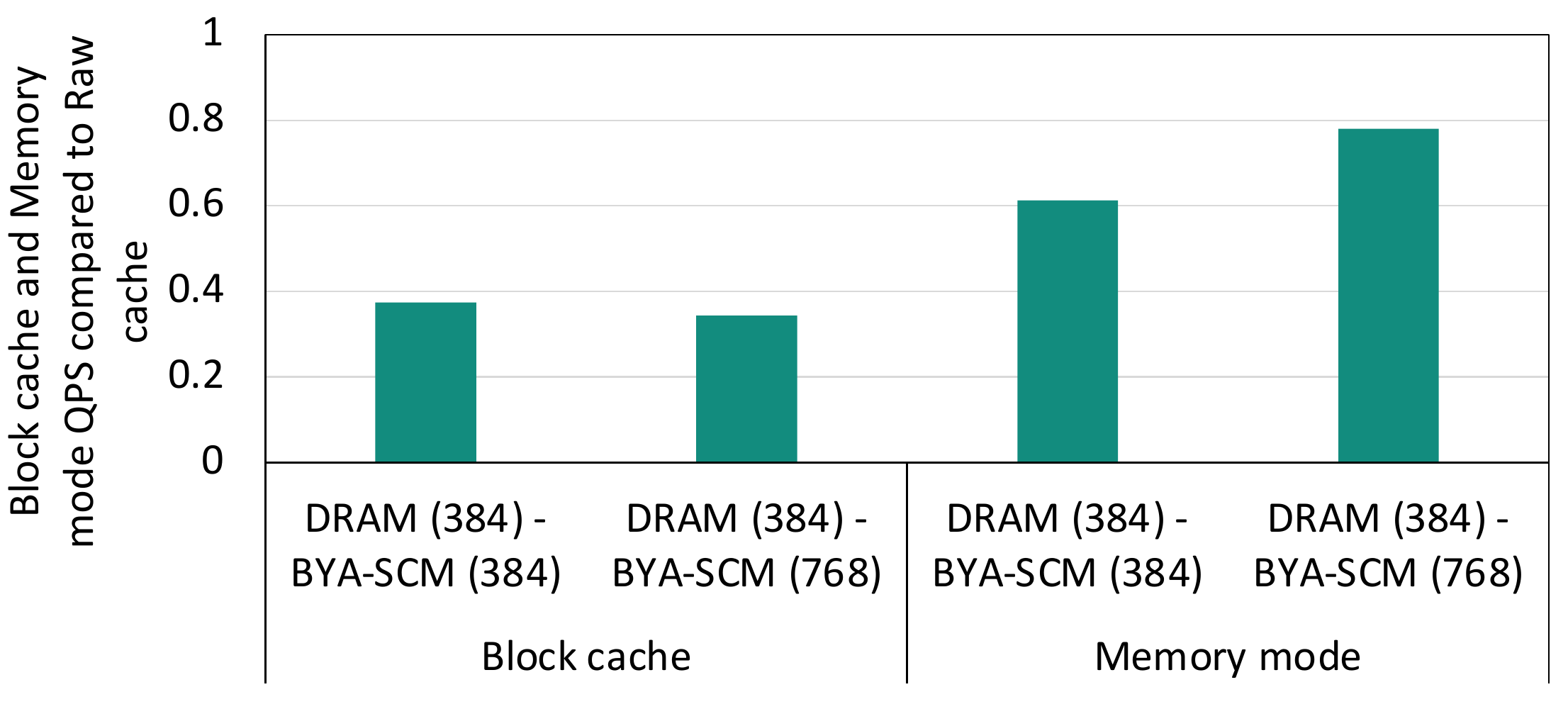} 
     \vspace{-2mm}
   \caption{\footnotesize{Caching efficiently comparison for \ifrmodel.}}
\label{caching}

\end{figure}
\subsection{Software design choices}
The first consideration of our software design is the efficient utilization of block-based storage. Block devices must always write a minimum of one block, even if only a single byte has changed. In order to efficiently utilize the drive, we use RocksDB \cite{rocksdb}. RocksDB is a key-value storage engine that provides low latency database operations. RocksDB's structure, such as efficient database sharding, allows for fast storage access. In addition, RocksDB uses an in-memory data structure for faster write operations. Since new writes go to memory (DRAM) first, RocksDB can compact many writes into a single large contiguous drive write. This significantly reduces SSD writes and increases SSD lifetime.

Second, due to the high temporal locality in embedding tables, caching effectively hides the latency of SSDs. However, to adequately utilize DRAM and \pmem, we must consider how we organize the cache. We first examine re-using the existing RocksDB block cache, which uses memory to cache data for fast read access, and store it in DRAM/pmem~ as shown in\cite{scm_block}. Another alternative is using the hardware-managed DRAM cache that comes with Intel Optane memory (memory-mode). In this mode, the hardware transparently uses DRAM as a direct-map cache of \pmem. We compared these two methods to a raw cache with access granularity equal to the embedding table row dimension. The raw cache hierarchically uses DRAM and \pmem~ in app-direct mode, where DRAM is the first level of cache, and \pmem~ is the second-level cache. Figure \ref{caching} compares the QPS of the raw cache versus both block cache and memory mode in two configurations for \ifrmodel. Since the block cache is designed for best performance in read-only cases, we did not find it performant in both hardware configurations (0.35-0.38$\times$ of the performance of the raw cache) for training DLRM models where the read/write mix is 50/50. This is because once an embedding row is updated, its location changes on disk due to write compaction and thus isn't accessible from the original block cache line. In addition, the block cache results in double-caching values, wasting capacity. Similarly, from Figure \ref{caching}, we can also see that memory mode is not helpful because of double caching. Hence, it is essential to design caching that exposes the unified capacity of DRAM and ~\pmem. For these reasons, we design an exclusive hierarchical cache using the app-direct mode to fully control access granularity and embedding table and row placement.

\begin{figure}[t]
    \centering

     \includegraphics[width=0.47\textwidth, trim={0.56cm 5.65cm 10.75cm 0.05cm },clip]{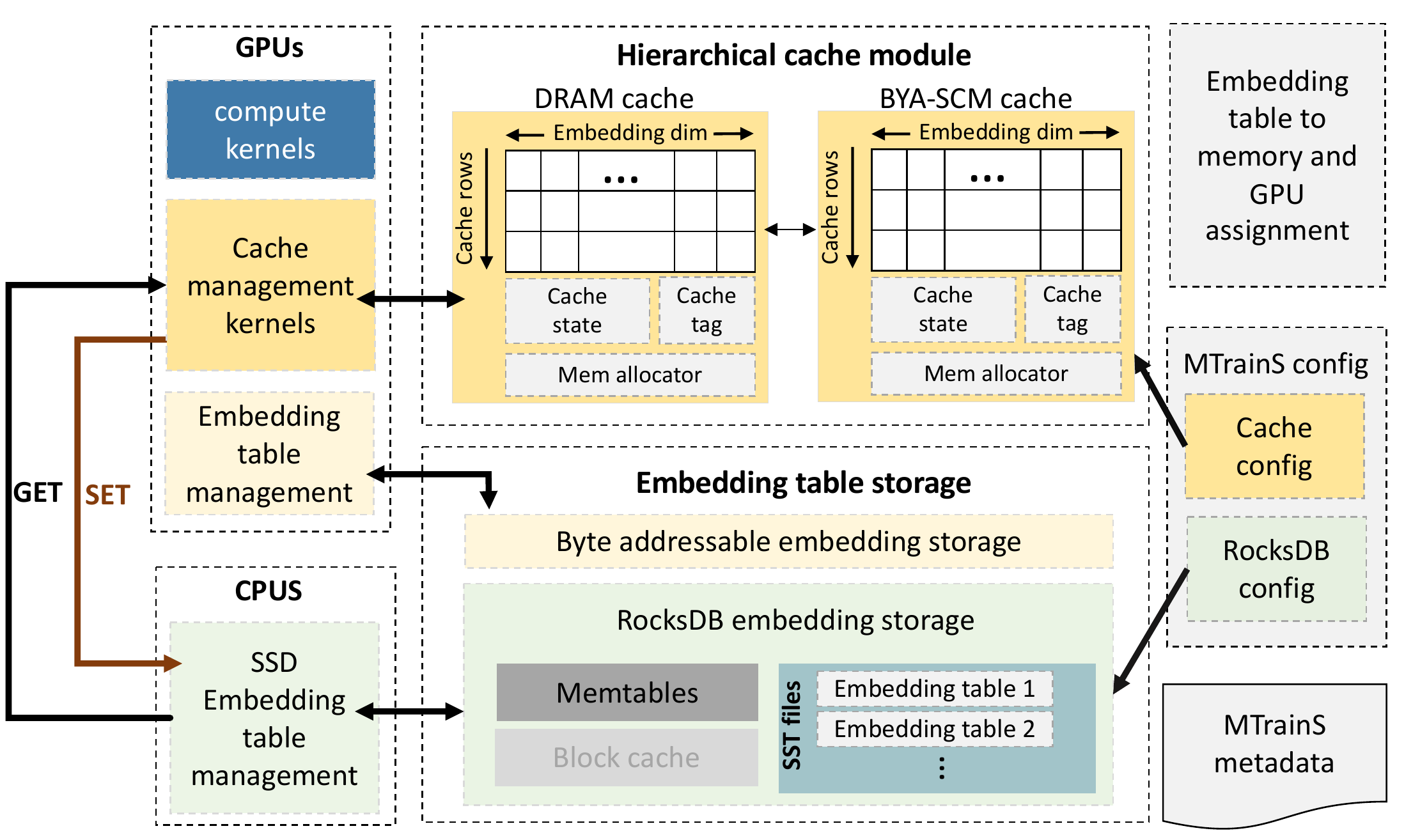}

\caption{\footnotesize{\ourway~ component and architecture.}}
        \label{figure:components}
\end{figure}

\section{\ourway~ design}\label{design}
\label{sec:design}
\subsection{Overview}
Adding heterogenous memories in DLRM system design requires considering the memory technologies' latency, size, and bandwidth differences and a closer look at our models, as discussed above. We design \ourway, an end-to-end training pipeline to leverage HBM, DRAM, \pmem, and \optane~or \flash~in DLRM. \ourway~ extends the embedding table storage to \optane~and/or \flash~ using \textbf{RocksDB-based embedding table storage}, giving us the flexibility to accommodate large TB scale models with varying sizes and BW embedding. Based on the temporal locality of our models (see Section \ref{section:locality}), \ourway~ implements \textbf{a hierarchical cache module} using the least recently used (LRU) policy for embedding tables stored in SSDs to hide the large SSD latencies. It uses DRAM and \pmem~ for caching and places hot embedding rows in DRAM for fast accesses and colder ones in \pmem, providing slower but ample cache space. Given the wide ranges of size and bandwidth requirements (see Section \ref{section:bw}) in embedding tables, to maximize the size and BW availability in the memories and storage, \ourway~ uses \textbf{embedding table placement} based on a mixed-integer linear programming solver, with table size and data volume per access (pooling factor) as inputs, and memory size and bandwidth as constraints, with the goal of increasing the bandwidth of embedding tables and minimizing tables' access time. Figure \ref{designoveriew} shows the overview of where \ourway~fits in DLRM, and Figure \ref{figure:components} shows the architecture of \ourway. 



\vspace{-2mm}
\subsection{Embedding table storage}
\ourway~ uses RocksDB-based storage to place embedding tables in block-addressable storage (\optane~ and \flash). Storing tables in SSDs allows for extensive storage per node for the big-size embedding tables. RocksDB\cite{rocksdb} is a key-value store optimized for high-speed storage. In the context of embedding tables, the key is the embedding table row index, and the value stored is the embedding row containing the weights. RocksDB organizes the key-value database into blocks (4KB in our implementation) in a Sorted String Table (SST) format. We sharded these databases (SST files) of embedding tables to load balance and for fast key lookup. Our RocksDB implementation uses memtable, located in DRAM, to optimize writes. However, as discussed above, we turn off the block cache because it does not benefit our DLRM workloads. For faster read, we use the MultiGet() \cite{multiget} API in RocksDB, optimized for batched lookups.  

\ourway~ also allows embedding table placement in the byte-addressable memories (HBM, DRAM, \pmem) using a two dimensional tensor. These tensors have the same structure in all memory types with different memory allocator parameters. 
\vspace{-2mm}
 

\subsection{Hierarchical cache module}
We leverage DRAM and \pmem~ as software-managed caches on top of the RocksDB embedding storage to hide the SSD's low bandwidth and high latency. The hierarchical cache module has a list of caches derived from a cache class and a configurable cache hierarchy that organizes the list of caches to multiple levels. Figure \ref{figure:components} shows the cache overview.
\subsubsection{Cache class} 
This class stores the hot rows of the embedding tables placed in SSDs. It has a 
\textbf{cache memory} that keeps raw embedding rows (as opposed to a block of multiple rows) as shown in Figure \ref{figure:components} because of the lack of spatial locality in the embedding tables. To allocate cache memory, the class has a \textbf{memory allocator} parameter that can be set to different memory types. The class also has tags and states. The \textbf{cache tags} track which Indices of the embedding rows reside in the cache and which cache entries are occupied/free. The \textbf{cache states} track each cache entry based on timestamp. DRAM and \pmem~ caches are an instance of the cache class with different cache memory allocators. 


\subsubsection{Cache hierarchy}
We use DRAM and \pmem~ as a configurable multi-level cache. The cache is configured as a one-level in the presence of only DRAM in the system. When we have DRAM and \pmem~ in the system, we organize the two memories as a two-level exclusive cache, with DRAM cache as the first level and \pmem~ cache as the second level. We use exclusive cache settings for more efficient use of the memory space. Note that we can use \pmem~ only as a one-level cache, but we didn't find this configuration performant because of its high latency. We identify cache location-specific operations, such as data movement between caches and which cache to access first, based on the cache hierarchy structure. Note that, we only have DRAM and \pmem~ cache here, but the cache structure can handle more than these two caches, such as multiple DRAM and SCM caches in a complex hierarchy.

\vspace{-2mm}
\subsection{Embedding table management}

This module accepts and responds to embedding table requests stored in all memory types. Embedding table lookups and updates are managed by the GPU for tables placed in the byte-addressable memories and by the CPU for tables placed in SSDs. While embedding lookups are initiated by the GPUs, they will be handed off to the CPU to get data from RocksDB embedding storage when Indices miss from the cache modules. Note that we can directly access the SSDs from the GPUs using GPU Direct Storage (GDS)~\cite{gds}, but we do not use it in our design to leverage the host side memory (DRAM and \pmem) as SSD caches. Using GDS limits the cache to HBM. This module exposes two APIs, GET and SET, from the SSD embedding storage to the rest of the trainer, as shown in Figure \ref{figure:components}. GET and SET APIs must synchronize the CPU and GPU to maintain data consistency. For efficiency, we designed a multithreaded management unit for RocksDB embedding storage in the CPU that looks up the RocksDB shards in parallel. The module is also responsible for initializing the embedding table weights before training starts. We provide two options for initialization: 
\vspace{-2mm}
\subsubsection{Pre-initialization}Initialize all the weights of the embedding tables stored in all memory types with random values following the desired distribution before training starts. 
\vspace{-2mm}

\subsubsection{Deferred initialization on read}
Embedding tables stored in SSDs have very large sizes. Pre-initializing all these weights takes a long time. We designed a deferred initialization technique to prevent this long initialization process at the start of training and to preserve SSD write endurance. In this technique, we initialize embedding values on-demand upon the first read if a key is not found inside the database Indices. During deferred initialization, to reduce initialization latency, we have a separate background thread that generates a queue of randomly initialized values following the desired distribution. This separate thread optimization is especially helpful in reducing latency when a single request tries to read many uninitialized rows. When we attempt to read an embedding row that has never been accessed, we consume values from the queue to randomly initialize the desired embedding row. This technique reduces writes by $\sim$15\% for \ifrmodel.

\vspace{-2mm}

\subsection{Cache management}
In our design, the cache is managed by the GPU, similar to the caching proposed in~\cite{yang2020mixedprecision}. We extend their work and design GPU kernels in FBGEMM\_GPU that support caches in multiple memories, such as DRAM and \pmem, in a hierarchy. GPU-managed cache gives us the advantage of using the higher GPU compute and BW capability to accelerate cache management operations. We discussed the cache management kernels below. 
\vspace{-2mm} 
\subsubsection{Tag/state lookup} 
This GPU kernel looks up the cache tags and states to check if the embedding rows are in the caches or the SSD storage for incoming embedding row requests. In a two-level cache, the kernel looks up the tags and states for both caches in parallel for efficiency. After determining hits and misses in all the caches, the kernel groups the Indices of incoming lookup requests based on the memory destination, i.e., DRAM, \pmem, or SSD.  
\vspace{-2mm} 
\subsubsection{Cache algorithm}
 The caching algorithm kernel uses the LRU policy. It looks up the grouped indices of each memory from the tag and state lookup kernel. Then based on this, 1) it updates the time of access (LRU status) of the cache hit indices in each cache state (DRAM and \pmem), 2) it determines the cache slots to insert the missed indices for the indices that are a miss,  3) it resolves the cache slots to evict if there are no free cache slots available. We explored the least recently (LRU) and least frequently (LFU) used caching algorithm. Our experiments on show that LRU provides $\sim$8-10\% better performance than LFU. This is because, while both LRU and LFU capture the temporal locality of embedding tables, with LRU after inserting embedding rows to the cache during the forward pass, the rows are still going to be the recently used row in the backward pass. This increases the chances of the rows being in the cache in the backward pass even when it is not the frequently accessed row, increasing cache hit rates in the backward pass compared to LFU.
\vspace{-2mm}
\subsubsection{Data lookup/update} It takes the grouped cache Indices from the tag lookup kernel and returns the rows. It also accepts new data or updates to Indices in the caches and returns evicted rows.
\vspace{-2mm}
\subsubsection{Data movement between DRAM, \pmem, and SSD}
In the one-level cache, if incoming embedding lookup requests are hits, the requested embeddings rows will be fetched by the requesting GPU. In case of misses, the control is passed to the CPU to access SSDs. The CPU then fetches the requested rows from the SSD storage. Newly fetched rows will be inserted into the first-level cache from SSD for fast access. If the cache is full, LRU rows will be evicted back to SSD to make room for the new rows. 

For a two-level cache, we access both DRAM and \pmem~ caches in parallel during lookup. Hits from both caches are returned to GPU. DRAM cache misses that are hits in \pmem~ cache and \pmem~ cache misses fetched from SSD are promoted to DRAM cache for fast access. When DRAM capacity is full, DRAM LRU rows are evicted to \pmem~ cache. Similarly, when \pmem~ cache is full, LRU rows are evicted back to SSD storage.

 

\begin{figure}
\centering
         \includegraphics[width=0.3\textwidth, trim={0.05cm 10.65cm 21.75cm 0.05cm },clip]{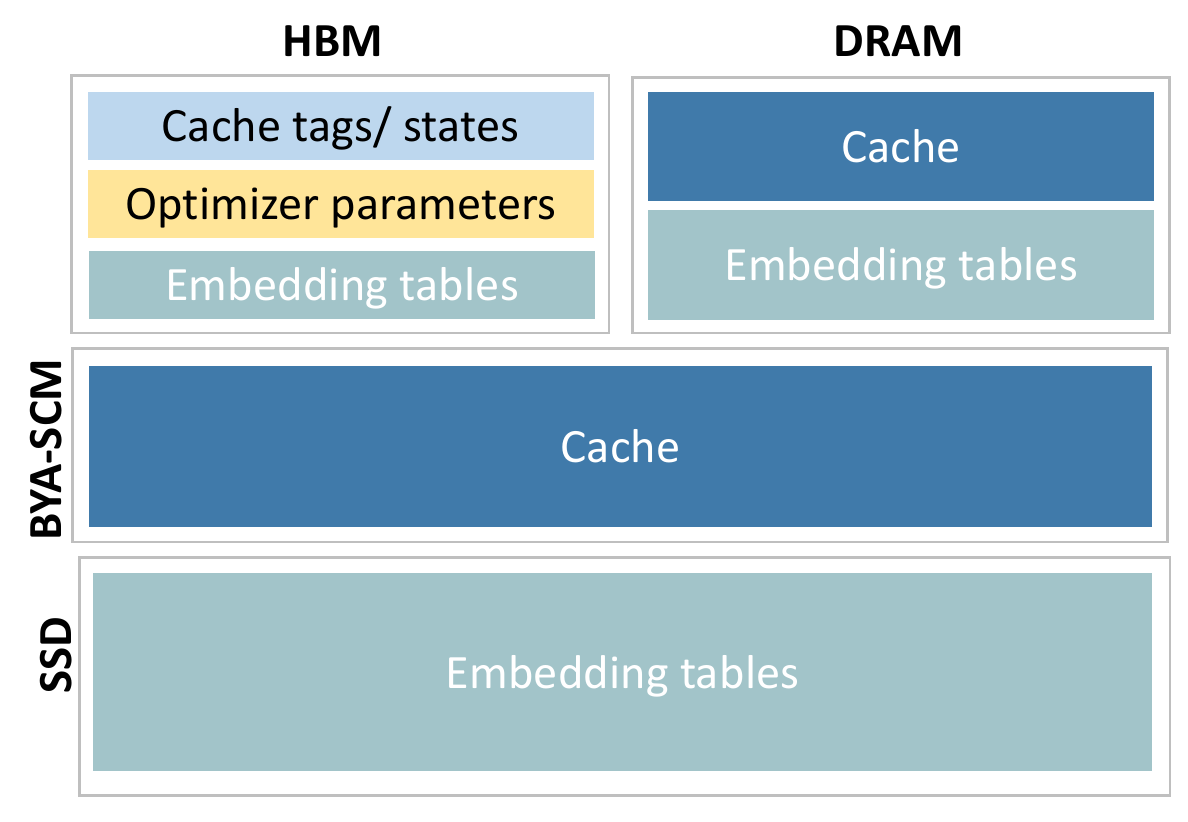}
\caption{\footnotesize{Memory allocation of \ourway.} }
        \label{figure:memplacement}
\end{figure}

\subsection{Memory and GPU assignment}\label{model assignement}
\subsubsection{Embedding table assignment}
A single DLRM model has various embedding tables having different sizes and BW. We have multiple memory types in our systems with varying sizes and bandwidths. While we use DRAM and \pmem~ for caching lower BW embedding tables stored in SSD, we can use HBM and some part of DRAM to place the high and medium BW embedding tables and still satisfy the latency and BW requirement of embedding tables stored in SSDs through caching. Figure \ref{figure:memplacement} shows memory allocation of MTrainS components. HBMs are used to store high bandwidth embedding tables, optimizer parameters, and the tags and states of the caches. DRAM is used to store medium BW embedding tables for caching hot embedding rows, and \pmem, for storing colder embedding rows of tables stored in SSDs. Our experiments show that using \pmem~ only for caching instead of an explicit assignment is better.
The search space for embedding tables assignment is vast because we can assign any tables in any of the memories. We use a simple mixed integer linear  solver to optimize the assignment. Table assignment is a complex problem, and it is possible that with a more sophisticated heuristic, we can achieve better table assignments than our current solution by considering other factors, such as locality. We leave such complex designs to future work.
\noindent\newline\textbf{Input variables:}
          The input variables for assignment are the sizes and BW of each embedding table in a model and the size and BW of the memory types.

\noindent\textbf{Constraints:}
The constraints for table assignment are:
\vspace{-1mm}
\begin{enumerate}
    \item Each table can only be assigned to one memory type, but each memory type can hold multiple tables. 
    \item The cumulative size of tables assigned to each memory type can not be larger than the memory size. 
\end{enumerate}
\vspace{-1mm}
\noindent\textbf{Objective function}:
   Minimize the total embedding lookup time, approximated according to Equation~\ref{eq:emb_lkup}. 
     \vspace{-1mm}
    
\vspace{-2mm}
\begin{equation}
\label{eq:emb_lkup}
\begin{split}
    lookup\_time = Max(time(g)), g~\epsilon~GPUCount \\
    time(g) = \sum_{M} \sum_{T_{g_m}} (D \times L \times p)/BW_{g_m}
\end{split}
\vspace{-3mm}
\end{equation}
\vspace{-2mm}

$M$ in Equation~\ref{eq:emb_lkup} stands for memory type (e.g. HBM or DRAM). $T_{g_m}$ represents embedding tables assigned to a specific memory type for a given GPU, and $BW_{g_m}$ represents BW for memory type $m$ for shard $g$. For example, for HBM, $BW_{g_m}$ represents HBM BW. For the shared DRAM, it would represent $DRAM\_BW/num_{gpus}$. We show the effect of placements in section \ref{pl}.
\vspace{-2mm}
\subsubsection{GPU assignment}Based on the table assignment, for $N$ GPUs, embedding tables assigned to the HBM of the $GPU_i$ will be handled by $GPU_i$. In addition, the embedding tables assigned to DRAM and SSD will be distributed to be managed by the $N$ GPUs by the table placement algorithm by minimizing lookup time in Equation~\ref{eq:emb_lkup}. 



\subsection{Pipelining} 
To hide some of the latency of accessing SSD for cache misses, we can pipeline access to the caches several batches in advance. Instead of sequentially 1) Fetch, 2) Preprocess, 3) Load on GPU, 4) Train, we split each step into its own stage and execute them simultaneously for different batches. In our case, we added a step: 4a) Prefetch Sparse Indices into cache before training. As long as we can maintain an invariant that embedding rows prefetched into the cache are not evicted until that batch has been trained, we can have an arbitrary number of batches in the pipeline. By adding additional stages between 4a) Prefetch and 4) Train, we can increase the latency hiding capability of the pipeline until it exceeds the typical SSD latency for a GET call. If the demanded bandwidth required to meet the QPS goals exceeds the capabilities of the SSD, no amount of extra stages will help.

\begin{figure}[t]

    \centering

     \includegraphics[width=0.5\textwidth, trim={0.05cm 11.55cm 1.55cm 0.05cm },clip]{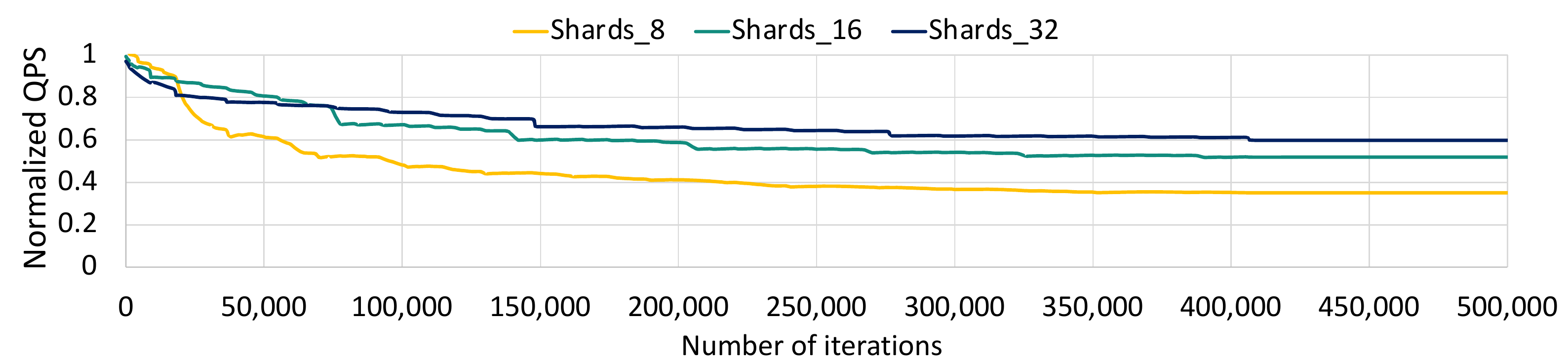}

\caption{\footnotesize{Impact of sharding RocksDB database on QPS}.}
        \label{figure:shards}
 
\end{figure}
\begin{figure}[!t]
    \centering

     \includegraphics[width=0.5\textwidth, trim={0.05cm 9.55cm 0.95cm 0.05cm },clip]{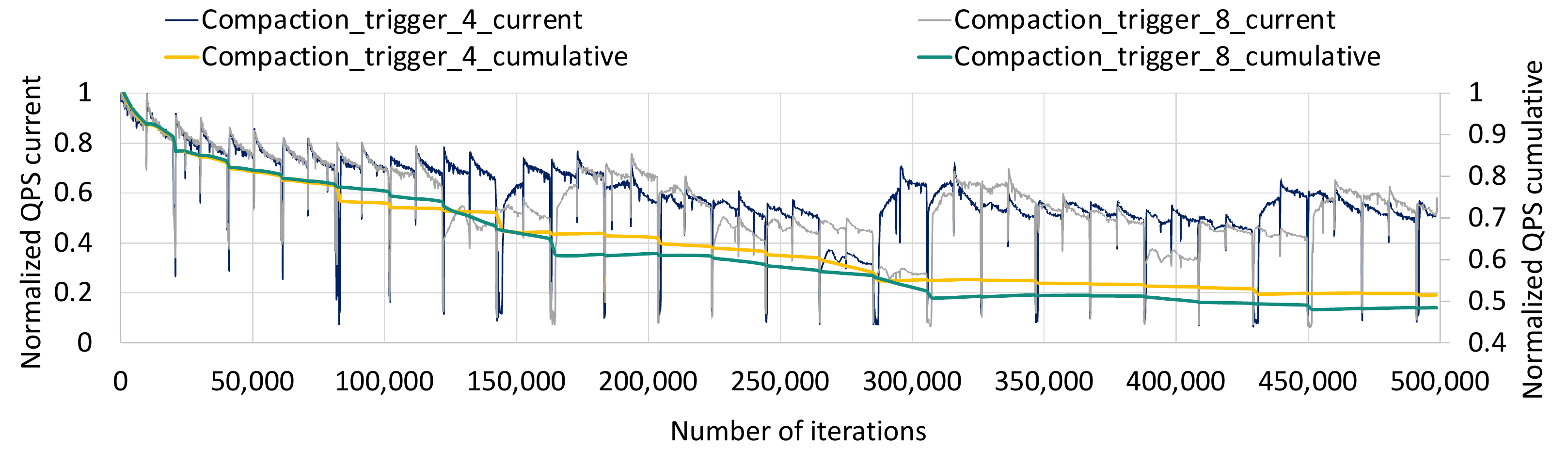}

\caption{\footnotesize{Impact of Database compaction on QPS.}}
        \label{figure:compaction}
\end{figure}
\subsection{\ourway~ configurations and metadata}
\subsubsection{\ourway~ metadata}The metadata keeps the memory and GPU assignment of all embedding tables and is used in every embedding table lookup to direct requests to the responsible memory.
\subsubsection{Cache config} This includes cache configurations to expose DRAM and \pmem~ memories hierarchically or to be used alone as the first layer of cache, cache row, and column sizes to fit the embedding dimensions of the target DLRM model. 

\subsubsection{RocksDB configs}
The RocksDB embedding storage exposes knobs to tune performance. The knobs include the number of CPU threads used for embedding table lookup, DB shards, compaction time, memtable sizes, and turning the block cache on/off. Sharding is one of the most important knobs, which increases key lookup efficiency and decreases compaction time. As seen in Figure \ref{figure:shards}, sharding DB increases QPS by up to 40\%. Another knob is database compaction, which is necessary for RocksDB to maintain a manageable database size during training. Synchronized database compaction from all RocksDB shards and trainers causes a major thundering herd problem that results in large memory and IO spikes. We observe considerable drops in QPS (over 50\% in some cases) during database compaction (seen in Figure \ref{figure:compaction}). Tuning compaction knobs, such as compaction trigger time as shown in Figure~\ref{figure:compaction}, improves the cumulative QPS by 5-8\%. In our experiments, we show the results for the best RocksDB configuration we found for each model.

\subsection{End-to-end trainer}
Figure \ref{splitmemflow} shows the end-to-end trainer. We first run the embedding table assignment and distribute embedding tables to HBM, DRAM, and SSD according to the optimal placement. Note that we don't need to run the placement for every training unless the model changes significantly. For the training data, the dense features are distributed across the batch dimension among multiple GPUs. The model embedding tables are distributed among the GPUs table-wise, and every GPU will handle the lookup for the embedding table assigned to it. For our current workloads, table-wise partitioning provides sufficient model parallelism across GPUs.
Then, during embedding lookups, \ourway~ will distribute the incoming input indices to the GPUs and memories. In Figure \ref{splitmemflow}, for example, Table 1 is placed in HBM and Table 2 and 3 in SSD. The GPU initiates embedding table lookup for the indices using \ourway~ in the forward pass. \ourway~ for example, gets index 1 of Table 2 from DRAM in the figure. Once the GPU gets the embedding rows from the memories, it performs aggregations and optimizers, then updates the weights in the respective memories in the backward pass.
\begin{figure}[t!]

\includegraphics[width=0.56\textwidth, trim={0.00cm 3.0cm 3.5cm 0.05cm },clip]{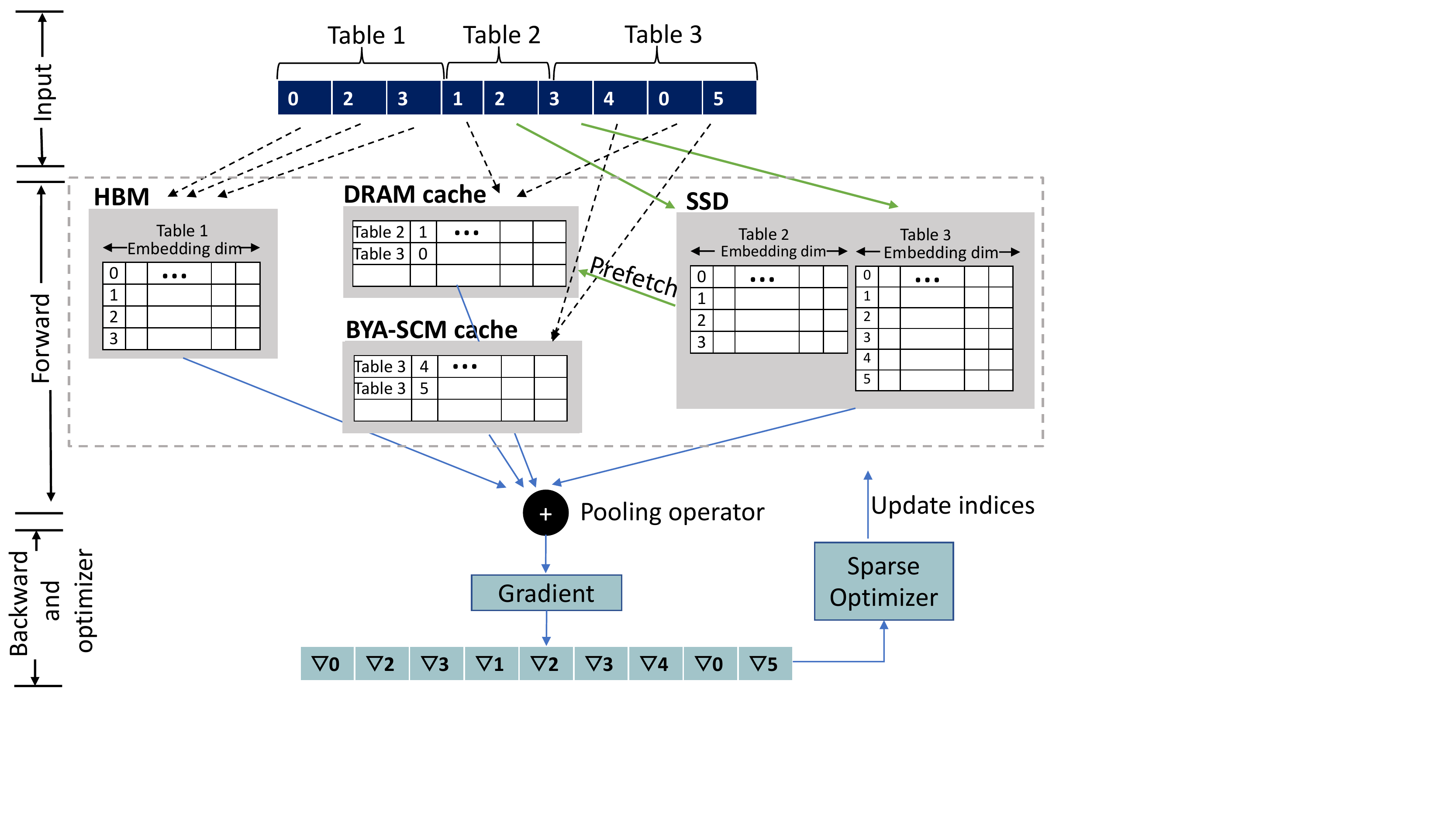}
 \caption{\footnotesize{\ourway~ trainer overview.}}
 \label{splitmemflow}
\end{figure}
 \section{Systems setup and implementations}\label{setup}

\subsection{Software design and implementation}
We use the PyTorch version of DLRM and implement a new EmbeddingBag that uses heterogeneous memories and storage for embedding table operations instead of using the default EmbeddingBag \cite{EmbeddingBag} implemented with PyTorch in DLRM. Our new EmbeddingBag is integrated with the FBGEMM\_GPU kernels we developed to manage cache and memory. Figure \ref{software} shows the software overview. 

In our experiments, we set up Intel DCPMMs in App Direct mode using IPMCTL \cite{ipcmtl}. We use Linux Kernel 5.4, which enables a volatile use of DCPMM. We then use NDCTL 6.7 \cite{ndctl} utilities to configure SCM in the devdax mode. This mode gives direct access to DCPMM, which is faster than filesystem-based access. We use DAXCTL \cite{ndctl}  to set up DCPMM in the system-ram mode so that DCPMM will be available in its own volatile memory NUMA node. To implement SCM in DLRM, we use the Memkind library \cite{memkind}. Memkind partitions the heap into multiple kinds of memories, such as DRAM and SCM, in the application space. We use MEM\_KIND\_DAX\_KMEM for SCM accesses. 

We use CUDA 11 for our GPU kernel implementations. As shown in the Figure, we utilize cudaMallocManaged \cite{CudaManagedMalloc} that uses a unified memory system to access HBM and DRAM from the GPU transparently. For \pmem, since the GPU can't access these types of memories directly with unified memory, we use cudaHostRegister \cite{CudaManagedMalloc}, which registers an existing host memory range already allocated by the Memkind library to CUDA. We have multiple cache management and computation kernels. Using PyTorch's torch.cuda.Stream \cite{stream}, we launch kernels that can run in parallel, such as looking up tags or getting data from DRAM and \pmem~ caches, in different streams. We then use torch.cuda.synchronize \cite{sync} to synchronize the kernels getting data from the cache before computation kernels start. While performing calculations for a batch, in parallel, we update tags/states and manage insertion, update, and eviction in all caches and SSD storage. Then ‌these kernels are synchronized with the subsequent batch data lookups to have updated cache/data for the next batch.
\vspace{-3mm}
\subsection{Workloads description and setup}
In our experiments, we use two of the most significant DLRM models derived from real use cases (\ifrmodel~and \adsmodel). These models show distinct features (capacity- and memory-bound) present in most of our models. We also use a model with 2x the size of \ifrmodel~ (\ifrmodelscaled), representing how the model will grow in the next two years. We show the characteristics of the models in Table \ref{tab:workloads}.    
\begin{figure}[t!]

\includegraphics[width=0.49\textwidth, trim={0.05cm 8.0cm 14.5cm 0.05cm },clip]{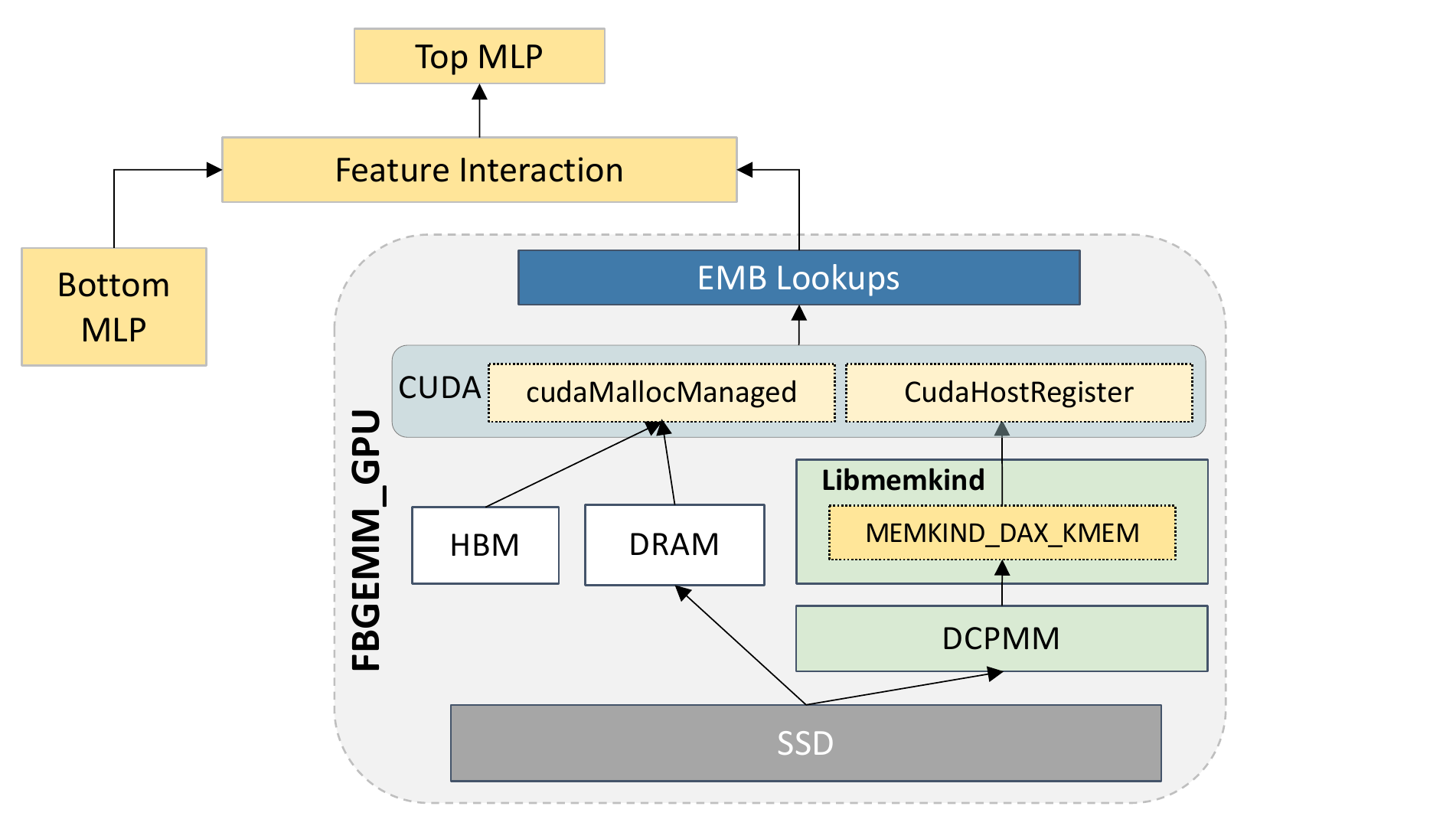}
 \caption{\footnotesize{\ourway~ software design.}}
 \vspace{-2mm}
 \label{software}
\end{figure}
\vspace{-2mm}
\subsection{Evaluation hardware description}
In our evaluations, we used an Intel Barlow Pass-based system, with 2 CPU sockets populated with Intel Ice Lake processors and  8 Nvidia A100 GPUs. 
This platform is designed for AI/ML, Deep Learning, and HPC applications, and has 8 NVidia A100 GPUs behind a fully connected NVLink / NVSwitch \cite{nvswitch} fabric. The hardware specification is shown in Table \ref{tab:spec}. 
\vspace{-2mm}

\begin{table}[t!]

\centering
\footnotesize
\begin{tabular}{|l|c|}
\hline
\textbf{Specification  }         & \textbf{System config }                      \\
\hline
OS /Linux kernel version                      & CentOS-8 /5.4.135                            \\
CPU model               &  Intel(R) Xeon(R) \\
&      Gold 6348 CPU @ 2.60GHz   \\
Sockets /Core per socket /Threads total                & 2 /28 /112 \\
L1I/L1D cache/L2/L3 cache            & 32 KB/48 KB/1.25 MB/42 MB                                \\
            
GPU model/ \#GPUs & A100-SXM4-40GB / 8\\
GPU Driver/CUDA Version &450.142.00/ CUDA 11 \\

HBM size& HBM2e 320 GB (40 X 8)\\
DRAM size               & DDR4 384 GB (32 GB X 12 DIMM slots) \\
SCM size                 & DDR-T 2 TB (128 GB X 16 DIMM slots) \\
\hline
\end{tabular}
\caption{System setup: hardware specs.}

\vspace{-2mm}
\label{tab:spec}
\end{table}
\begin{table}[t!]
\centering
\footnotesize

\begin{tabular}{|l|c|c|c|c|c|}
\hline
Config. & HBM & DRAM & \pmem & \optane & Nand \\
&&&&& SSD             \\
\hline
\basecfg & 320 & 384 & & & \\
\sysOne & 320  & 384 & & & 8192 \\
\sysTwo & 320  & 384 & & 2048& \\
\sysThree & 320  &  384 &384&  & 8192\\
\sysFour & 320  &  384 &768 &   & 8192\\
\sysFive & 320  & 384 & 384 & 2048 & \\

\hline
\end{tabular}

\caption{\footnotesize{System Configurations (all sizes in GB).}}
\label{tab:systems}
\end{table}
\subsection{Server design}
We consider diverse server designs by varying the memory and storage types and sizes. Table~\ref{tab:systems} summarizes the configurations we used in our experiments. In all of our system configurations, we limited the DRAM size to 384GB. We use two sizes of \pmem, 384GB and 768GB. We chose these sizes because we want to determine the ratio of DRAM and \pmem~required in the system. We also experiment with other configurations, such as increasing the \pmem~ size. However, increasing \pmem~beyond 768GB does not show additional benefit for our existing workloads because we will be compute-bound at this config. Similarly, with \optane~ with 768GB configuration.      

We use half of the DRAM in our system (192 GB) for caching and the rest to store smaller size embedding tables and for other system requirements of DLRM. We use all of BYA-SCM for caching, i.e., 360GB in 384GB configurations and 720GB in 768GB configurations. The remaining BYA-SCM is used for optane metadata. 


\section{Evaluation}\label{experiments}
\subsection{Baseline}
We compare \ourway~ to the baseline system configuration shown in Table \ref{tab:systems}. Our implementation in the baseline also uses an integration of DLRM and FBGEMM\_GPU. We use HBM and DRAM in the baseline for embedding storage and caching. This implementation uses the same techniques as CDLRM \cite{balasubramanian2021cdlrm}. While CDLRM uses the CPU to manage the cache, for a fair comparison with \ourway, which uses GPU for fast cache management, we move the cache management in CDLRM to GPU. We call this CDLRM+. We also added the efficient embedding table placement in CDLRM+. Our experiments compare the number of nodes required to train a model running CDLRM+ with the performance we get with a single node running \ourway. To compare performance, we use target QPS, an acceptable QPS in our data center, to train a specific model based on how often we need to train models and the rate at which new training data becomes available. QPS in our experiment represents the number of input data we can use to train a model per second, including the forward and backward pass. It is inversely proportional to the training time.   

\subsection{Training efficiency}
One of the principal arguments for adopting denser memory technologies is to use fewer nodes for specific model training and improve power and cost-efficiency. Hence, we first compare the overall deployment efficiency of \ourway~and CDLRM+ for \ifrmodel~and \adsmodel~ based on the target QPS.
\begin{figure}[]

 \centering
 \includegraphics[width=0.43\textwidth, trim={0.25cm 6.8cm 5.3cm 0.05cm },clip]{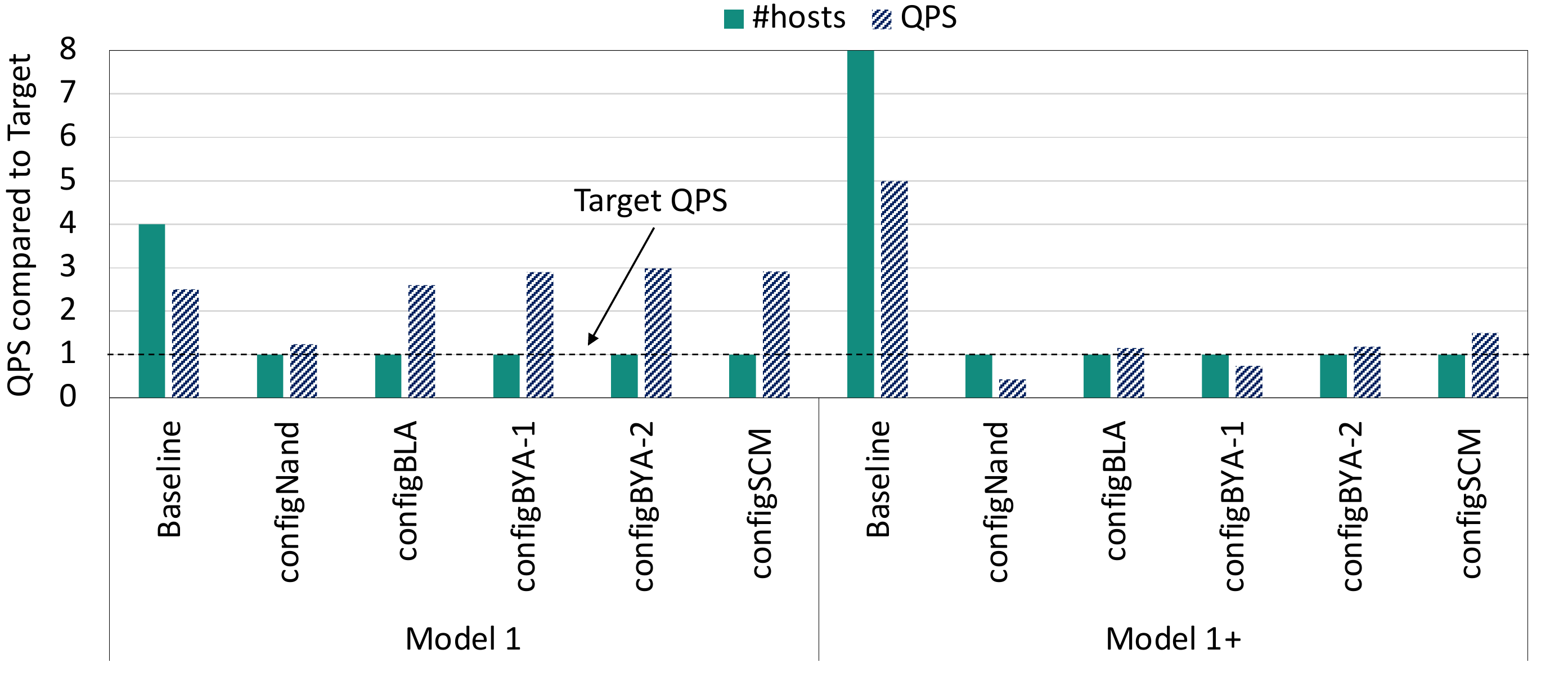}
 \caption{\footnotesize{ Training QPS and number of host Comparison of CDLRM+ and \ourway~ with SLA QPS target as the baseline for \ifrmodel~ and \ifrmodelscaled~}}
 \label{scale_model_1}
\end{figure}
\begin{figure}[]

 \centering
\includegraphics[width=0.44\textwidth, trim={0.25cm 7.1cm 5.3cm 0.05cm },clip]{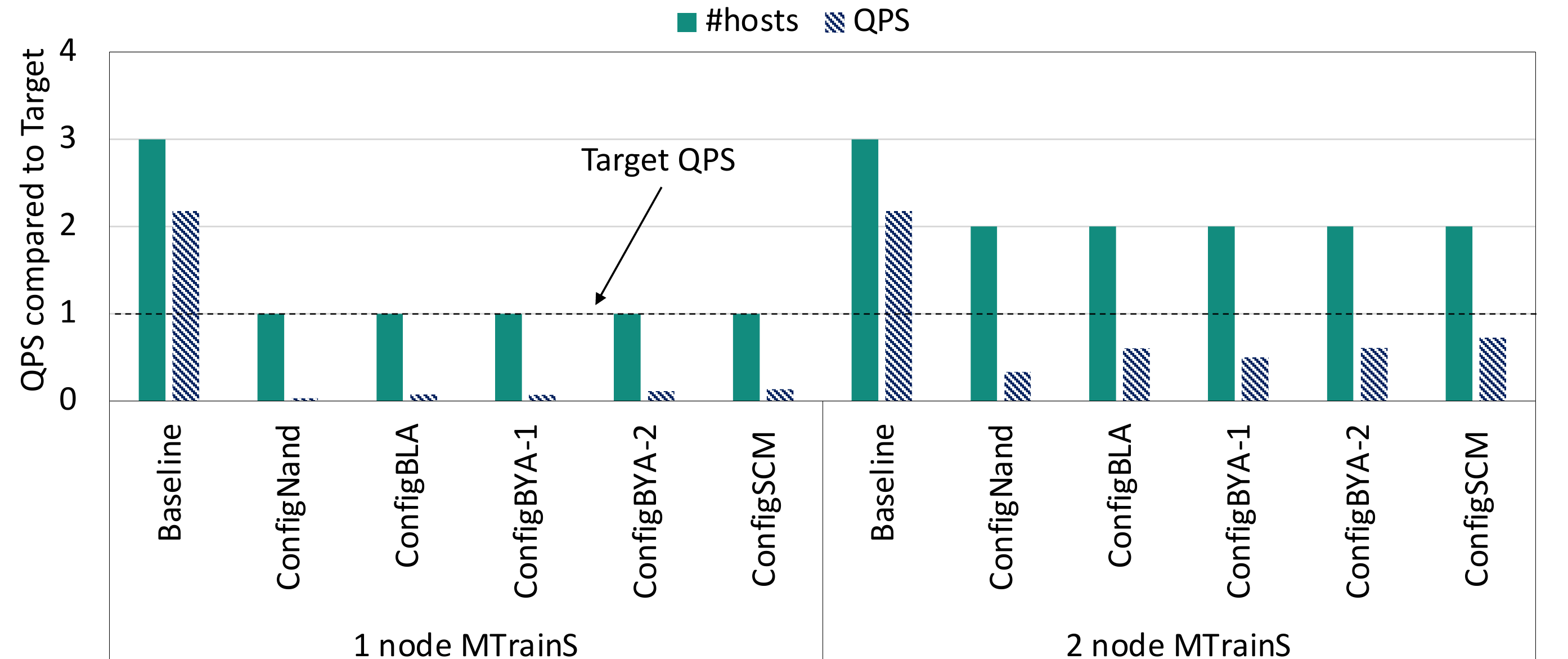} 
 \caption{\footnotesize{ Training QPS and number of host Comparison of CDLRM+ and \ourway~ with SLA QPS target as the baseline for \adsmodel.}}
 \label{scale_model_2}
\end{figure}

Figure \ref{scale_model_1} shows the efficiency comparison of \ifrmodel~and \ifrmodelscaled, which are memory capacity bound. Note that \ifrmodelscaled~ is the future scaling of \ifrmodel~ with a larger size (2$\times$). These models require 4-8 hosts to load and train with the baseline CDLRM+. With \ourway, one host can provide sufficient extended memory capacity to load and train the model. We can also reach the target QPS (SLA) in such a setting, as seen in Figure \ref{scale_model_1}. While the baseline without \ourway~ achieves higher QPS, it requires 4-8$\times$ more hosts, so this performance is stranded and does not contribute additional efficiency to the deployment because we already met our training QPS requirement. As a result, \ourway~ can improve power and cost efficiency for \ifrmodel~ and \ifrmodelscaled~ while providing SLA performance. Our experiments show that models scaled beyond \ifrmodelscaled~require multi-node training, each node leveraging \ourway. Nevertheless, \ourway~ still provides a lower host count because it can store a larger model per host compared to the baseline.  

Figure \ref{scale_model_2} shows a comparison of \ourway~vs CDLRM+ for \adsmodel~ (BW bound workload). This model requires 3 hosts to load and train with the baseline configuration using CDLRM+. While \ourway~allows the model to be loaded and trained on one host, it does not meet the QPS requirements because of the higher memory bandwidth required by the model. Also shown in the figure, using 2 nodes, each running \ourway, significantly improves the performance of \adsmodel~ ~ compared to the single node. Regardless, the 2 nodes still fail to meet the QPS target because the additional BW with SCMs in the system, even with 2 nodes, is not enough to accommodate the high BW demand of \adsmodel. However, such capability extends the efficiency of research and development without high QPS production requirements.




Next, we evaluate the performance of the various \ourway~ system configurations compared to \flash~
to understand the performance implication of \pmem~ and \optane~ compared to \flash. In Figure \ref{fig:qps_ifr} and \ref{fig:qps_ifr_scaled}, we compare QPS achieved for \ifrmodel~ and \ifrmodelscaled~ for different configurations of \ourway. As shown in Figure \ref{fig:qps_ifr}, for \ifrmodel, using \optane~(\sysTwo) instead of \flash~ (\sysOne) increases QPS by 2$\times$. Using \pmem~with \flash~(\sysThree~ and \sysFour) increases cache size. Hence lowers traffic to \flash~ and therefore provides 2.4$\times$ QPS. Our results indicate no further improvement in QPS for the setup with both \pmem~and \optane~ for \sysFive ~ because once we increase the cache size per host using \pmem, the access to SSDs is out of the critical path.

\begin{figure}[t!]

  \begin{subfigure}{.50\columnwidth}

\includegraphics[width=1.16\textwidth, trim={0.6cm 11.5cm 16.9cm 0.05cm },clip]{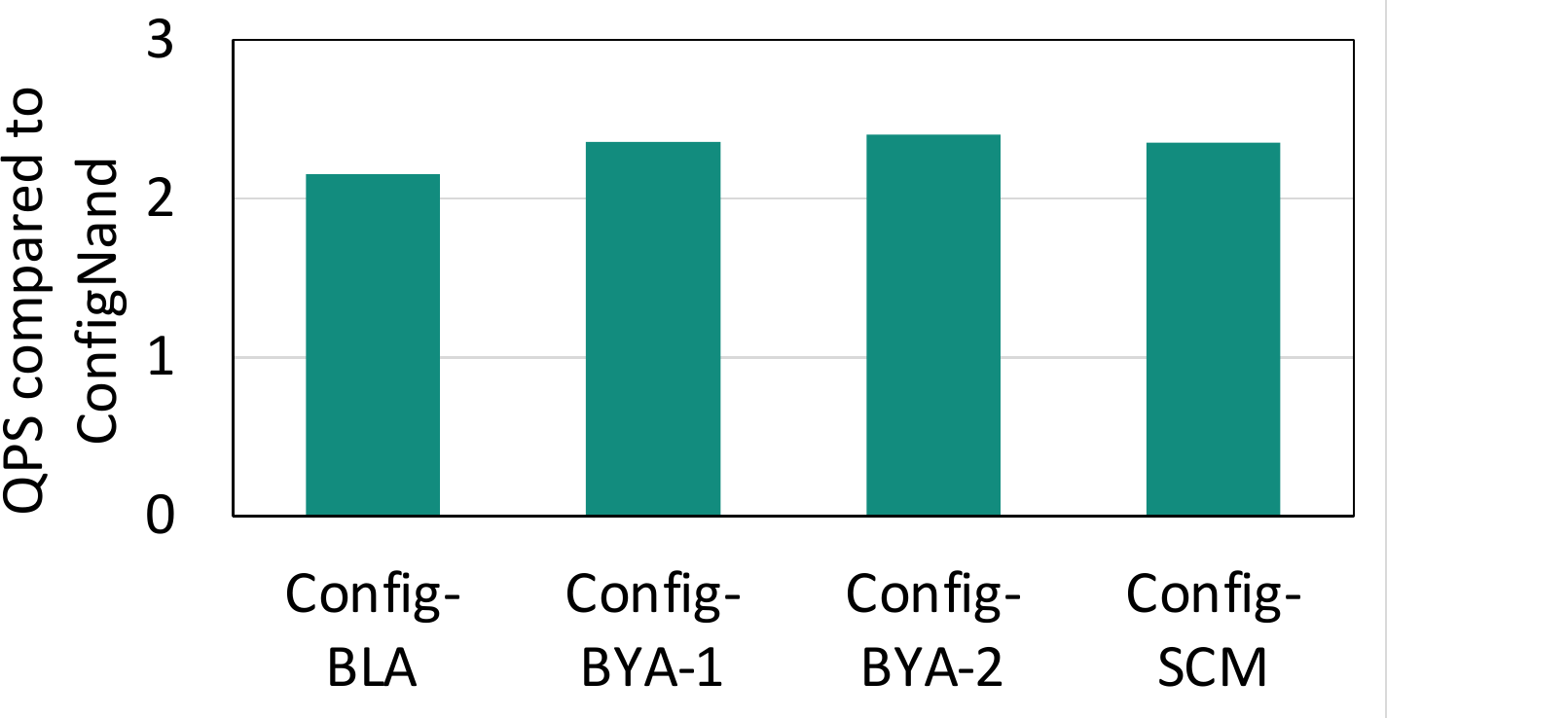}
\caption{\textmd{\ifrmodel} }
\label{fig:qps_ifr}
    \end{subfigure}
\begin{subfigure}{0.46\columnwidth}
\includegraphics[width=1.14\textwidth, trim={0.1cm 11.5cm 18.9cm 0.05cm },clip]{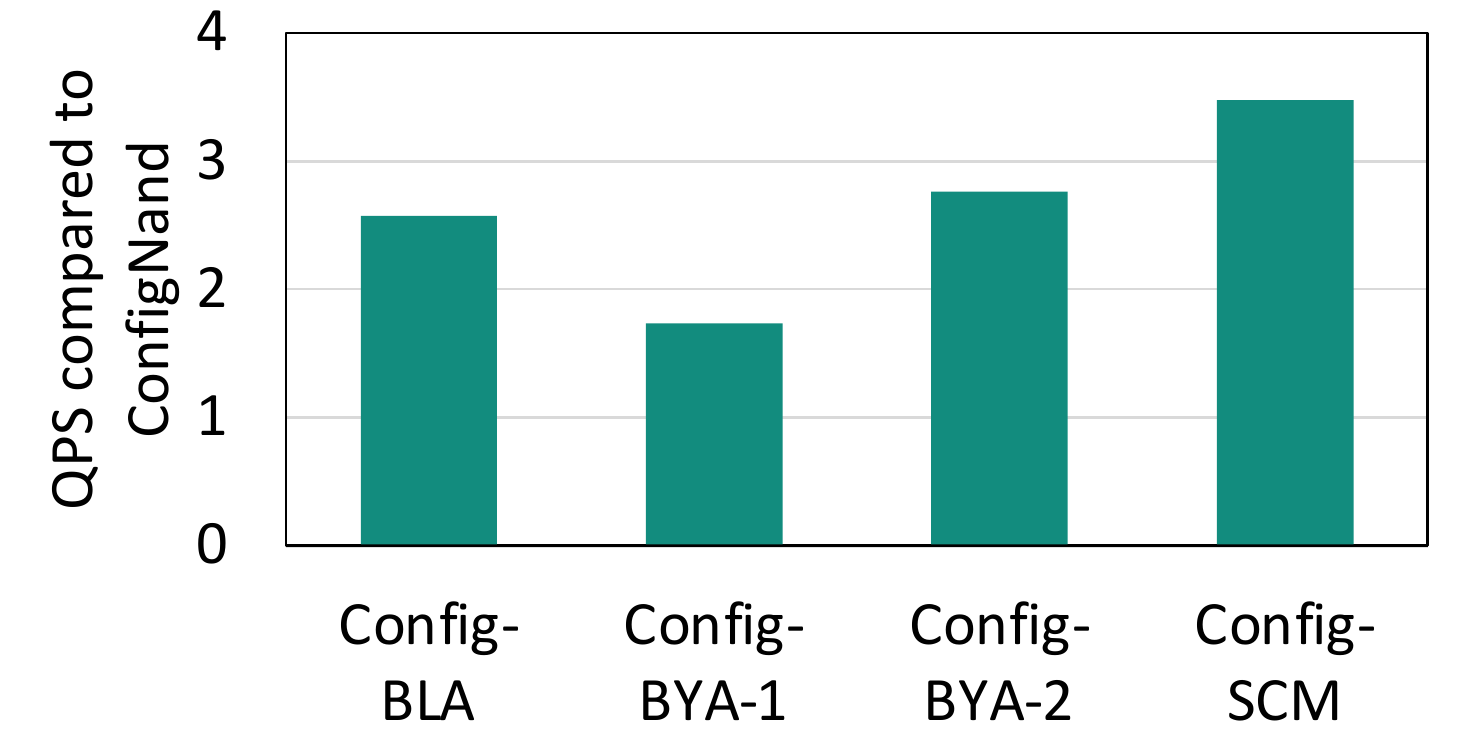}

\caption{\textmd{\ifrmodelscaled}}
\label{fig:qps_ifr_scaled}

    \end{subfigure}
 \caption{\footnotesize{QPS comparison of different configuration of \ourway~ with \flash~ as the baseline for \ifrmodel~and \ifrmodelscaled.}}
\end{figure}

\begin{figure}[t!]

  \begin{subfigure}{.50\columnwidth}

\includegraphics[width=1.21\textwidth, trim={0.25cm 11.5cm 16.5cm 0.05cm },clip]{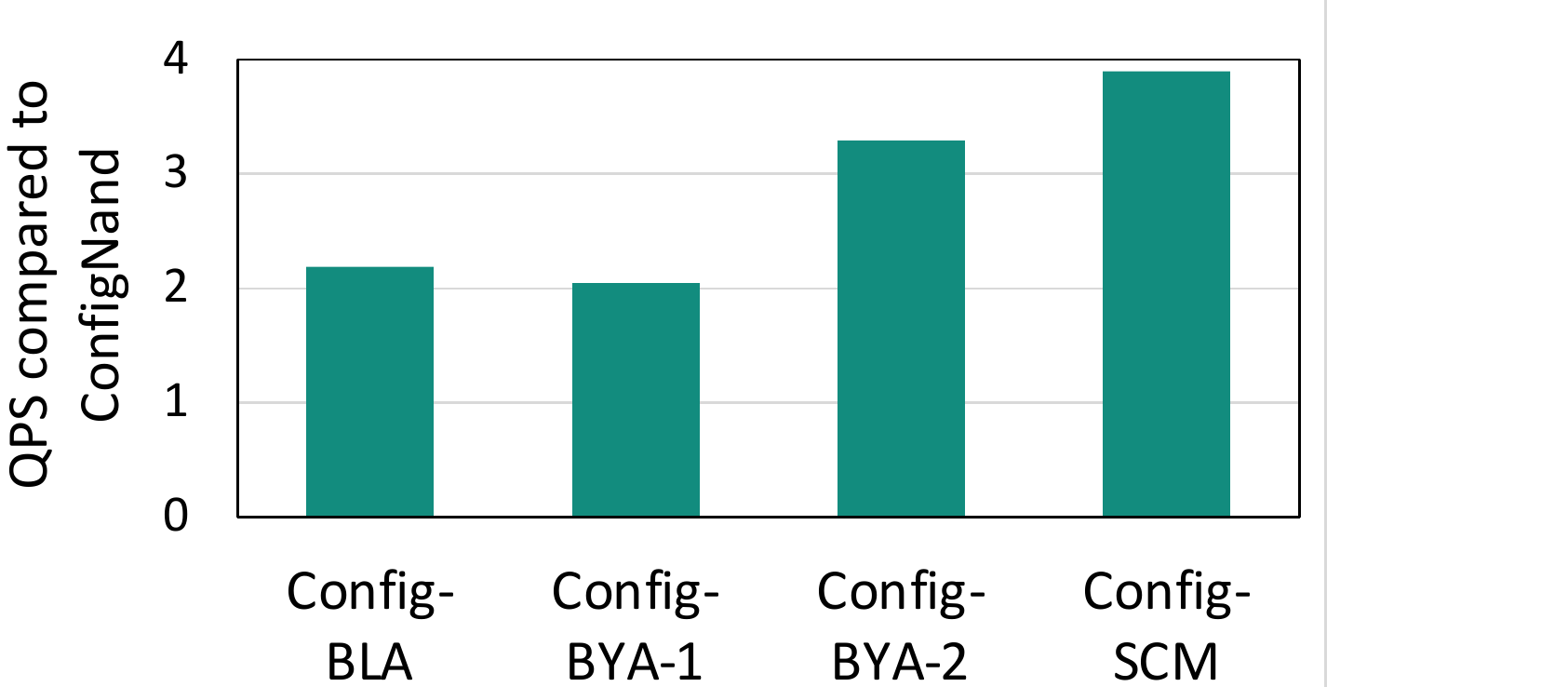}
\caption{\textmd{1 Node \ourway} }
\label{fig:qps_ads}
    \end{subfigure}
\begin{subfigure}{0.46\columnwidth}
\includegraphics[width=1.14\textwidth, trim={0.1cm 11.7cm 19.09cm 0.05cm },clip]{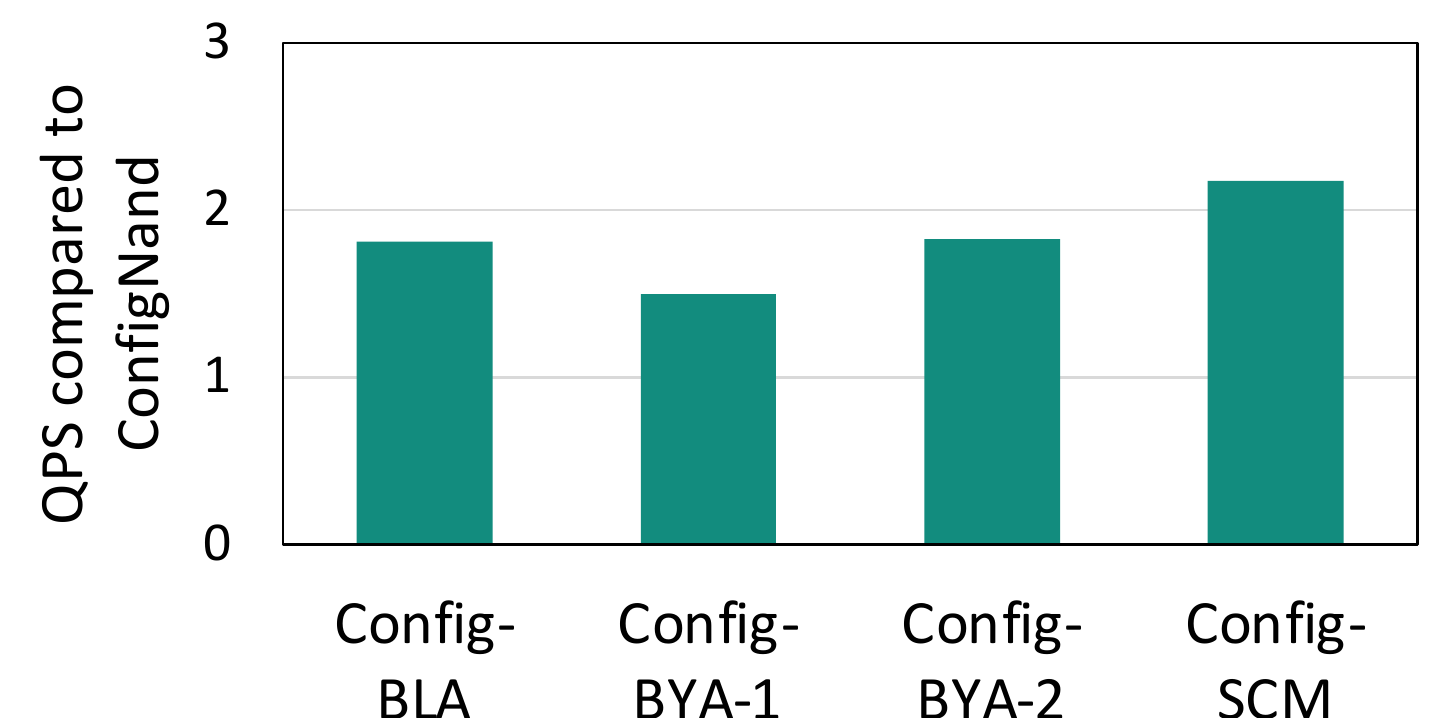}

\caption{\textmd{2 Nodes \ourway}}
\label{fig:qps_ads_2}

    \end{subfigure}
 \caption{\footnotesize{QPS comparison of different configuration of \ourway~ with \flash~ as the baseline for \adsmodel.}}
\vspace{-6mm}
\end{figure}

For \ifrmodelscaled, shown in Figure \ref{fig:qps_ifr_scaled}, \flash~with 384GB of \pmem~(\sysThree) increases QPS by 1.73$\times$ whereas, 768GB of \pmem~(\sysFour) increases by 2.76$\times$ due to the increased cache hit rate. When we use both \pmem~with \optane~(\sysFive), we achieve 3$\times$ QPS. \ifrmodelscaled~has higher BW requirement due to increased embedding dimension. With \flash~we already read 4KB blocks per IO, which is larger than the embedding dimension. Nevertheless, increased embedding dimension results in fewer embedding rows maintained in the cache (DRAM or \pmem), given fixed cache size. This increases the miss rate and hence results in increased IO to the SSDs. \optane~with higher IOPS can support such an increase in IO, hence showing higher QPS. 

In Figure \ref{fig:qps_ads}, for \adsmodel~ with 1 node \ourway, because of \optane's higher bandwidth and lower latency compared to \flash~ in \sysOne, performance increases by ~2.2$\times$. By adding \pmem~ we get 2$\times$ and 3.2$\times$ more performance because the cache hit increases. Similar to \ifrmodel, the combination of \pmem~ and \optane~ achieves the best performance. In figure \ref{fig:qps_ads_2}, the model is sharded between 2 nodes, and the model size per node is half of 1 node \ourway. The performance improvement for 2 nodes follows the same pattern as 1 node \ourway, but the speedup drops compared to 1 node because SSD traffic decreases when the model size per node is reduced, hence caching has less impact compared to 1 node. Although compared to \sysOne, the various configurations show performance improvement for \adsmodel, using SCMs does not provide adequate performance (QPS target) for memory bandwidth-bound workloads, as seen in Figure \ref{scale_model_2}.
\begin{figure}[t!]
\begin{subfigure}{.48\columnwidth}
\includegraphics[width=1.3\textwidth, trim={0.05cm 9.0cm 14.5cm 0.95cm },clip]{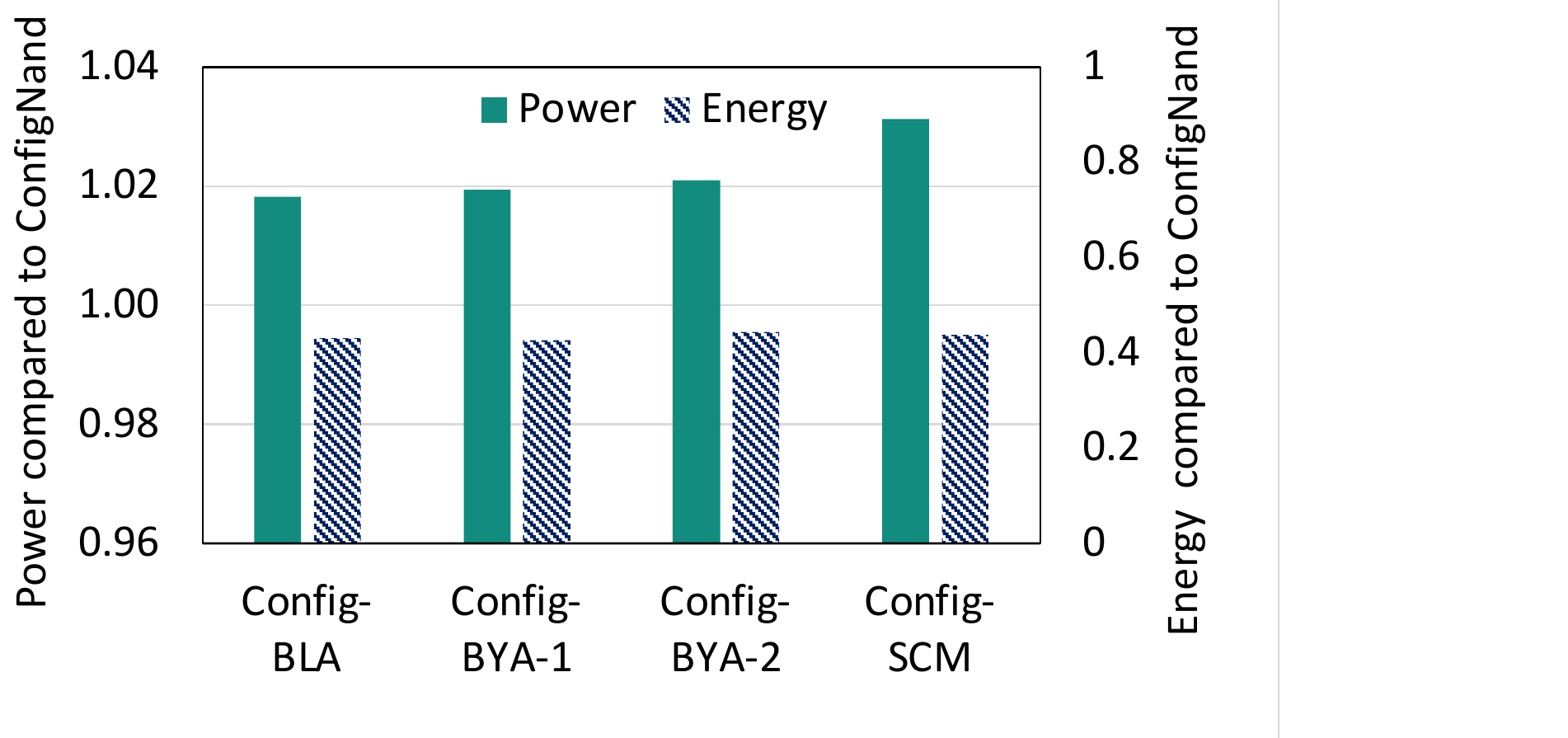}
\caption{\textmd{\ifrmodel} }
\label{figures:power_1x}
    \end{subfigure}
\begin{subfigure}{0.48\columnwidth}
\includegraphics[width=1.0\textwidth, trim={0.05cm 8.6cm 18.9cm 0.95cm },clip]{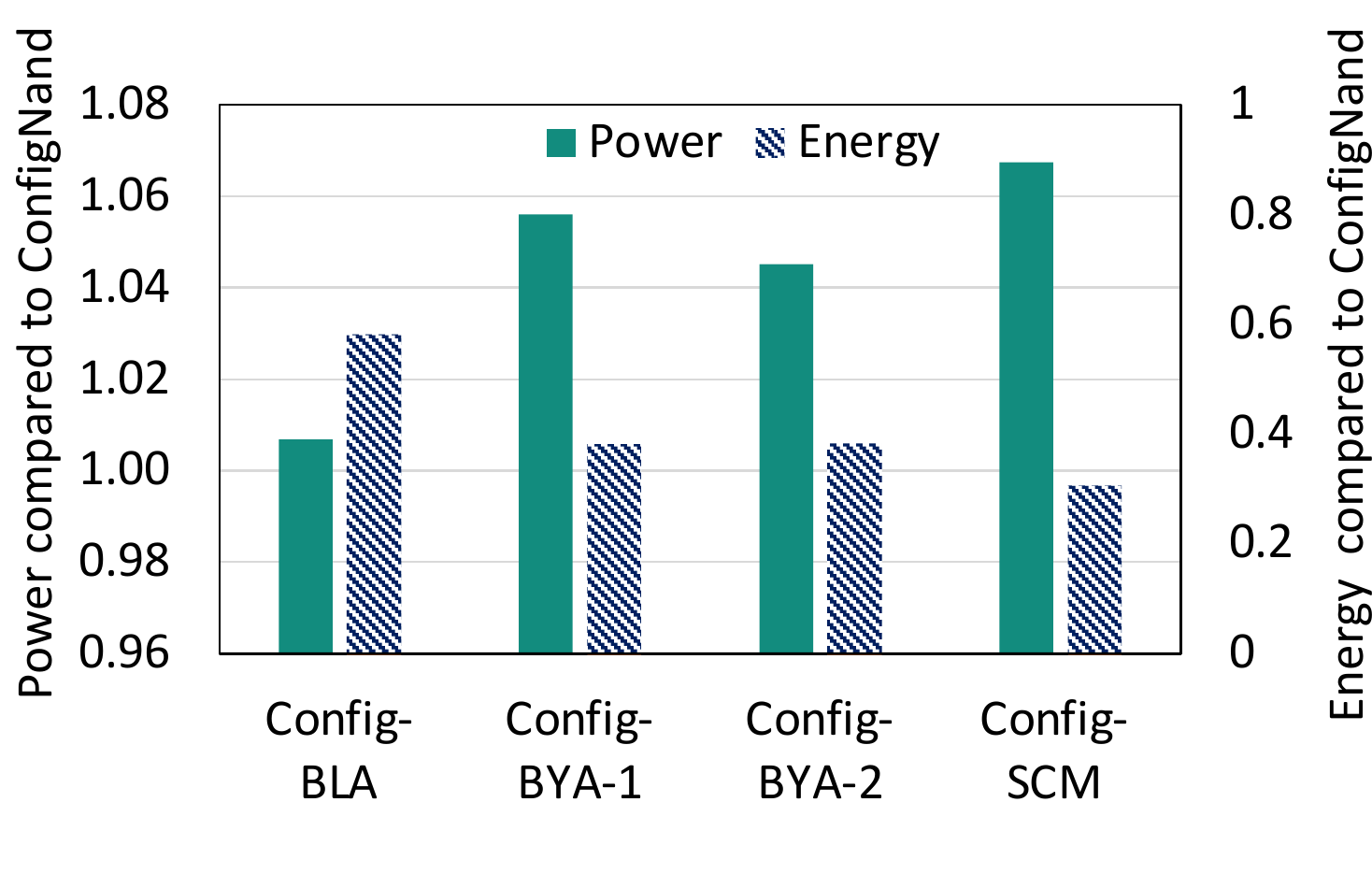}
\caption{\textmd{\ifrmodelscaled}}
\label{figures:power_2x}
    \end{subfigure}
\caption{\footnotesize{
Average power and total energy consumption.}}

\end{figure}


\begin{figure}[t!]
\begin{subfigure}{.48\columnwidth}
\includegraphics[width=1.1\textwidth, trim={0.05cm 10.0cm 18.5cm 0.05cm },clip]{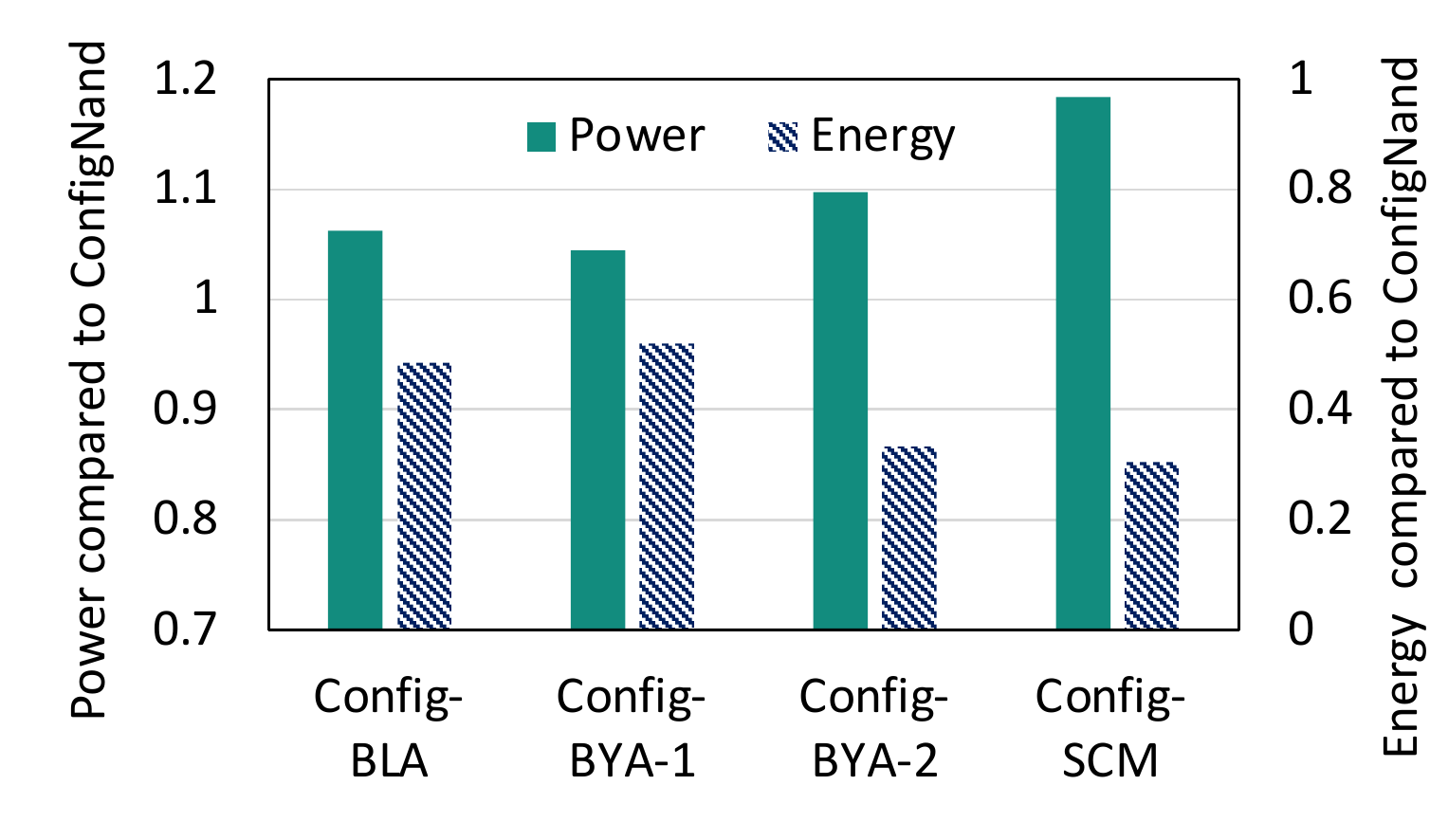}
\vspace{-7mm}
\caption{\textmd{1 node \ourway} }
\vspace{-1mm}
\label{figures:power_ads}
    \end{subfigure}
\begin{subfigure}{0.48\columnwidth}
\includegraphics[width=1.05\textwidth, trim={0.05cm 10.0cm 18.5cm 0.05cm },clip]{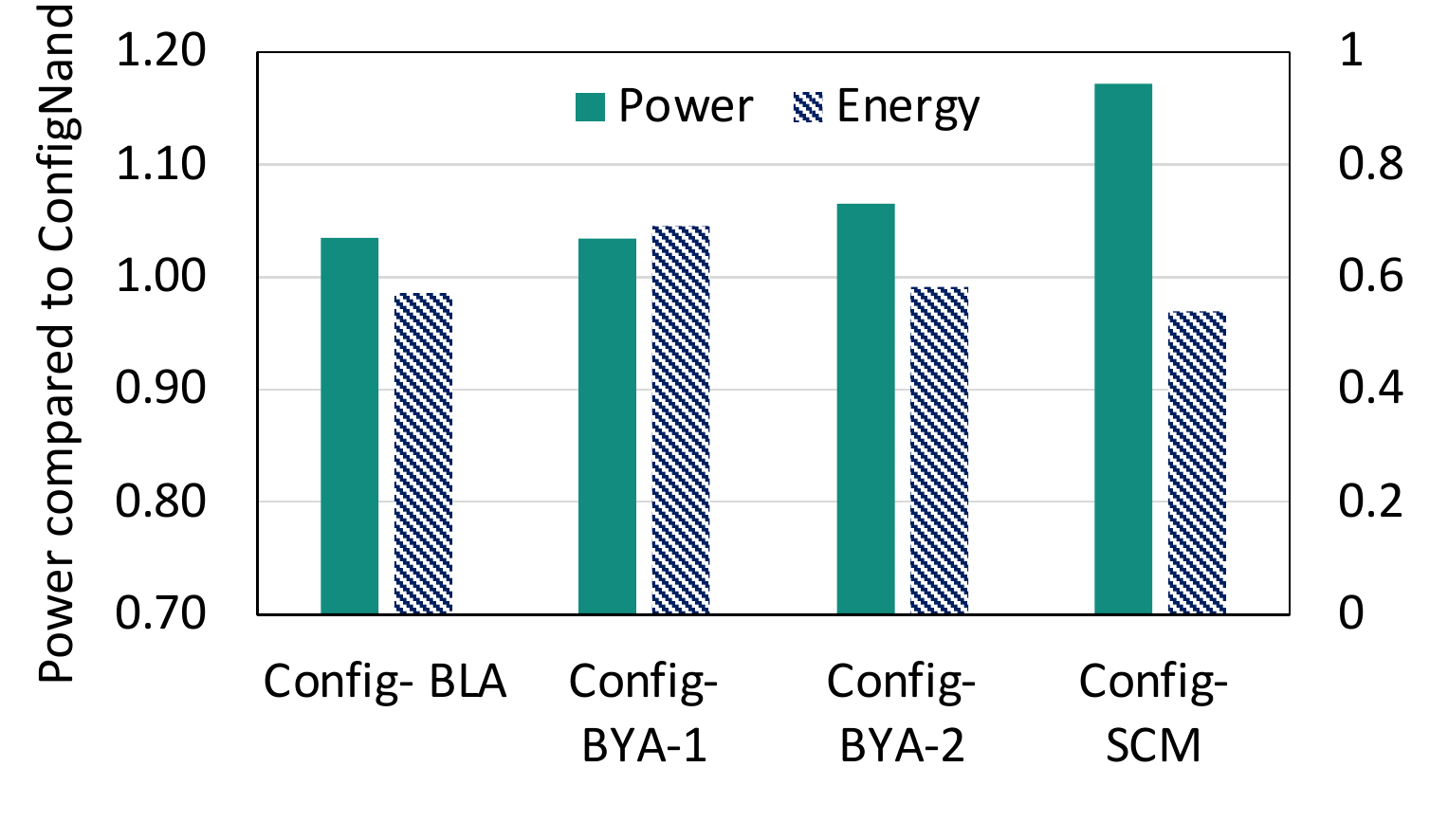}
\vspace{-7mm}
\caption{\textmd{2 nodes \ourway}}
\vspace{-1mm}
\label{figures:power_ads_2}
    \end{subfigure}
    \vspace{2mm}
\caption{\footnotesize{
Average power and total energy consumption.}}
\vspace{-3mm}

\end{figure}

\vspace{-1mm}
\subsection{\textbf{Power and energy analysis}}

 In Figure ~\ref{figures:power_1x} and Figure~\ref{figures:power_2x} we show the power and energy of the various \ourway~ configurations for \ifrmodel~ and \ifrmodelscaled~ to study the increase in power and the total energy usage when introducing \pmem~ and \optane~ to our system. 
The power consumption of adding \pmem~ and \optane~ only increases the overall platform power consumption by 1-3.2\%. This is because of the low power consumption of these individual units and because major power consumption contributors are the GPU, CPU, and DRAM. This extra power consumption per node is justified when considering reduced execution time and overall Energy consumption ($Energy = Power \times Time$). We observe a 60\% to 70\% reduction in the energy consumption across the two models compared to \sysOne. Figure \ref{figures:power_ads} and \ref{figures:power_ads_2} show the power and energy of \ourway~ for \adsmodel. In this case, adding SCMs increases the power by 3-18\%. The higher power in \adsmodel~ is because there are more caching and embedding storage operations in \adsmodel~ due to the larger data access volume compared to \ifrmodel~that increases the power of \pmem~ and \optane.

Figure \ref{fig:baseline_energy} shows the power and energy of \ifrmodel~ and \ifrmodelscaled~  using \ourway~ compared to the baseline configuration with only HBM and DRAM. Compared to training using the baseline, which requires 4-8 nodes to accommodate the models, we see $\sim$ 1/4 - 1/8  power reduction with \ourway. This reduction is mainly driven by the decrease in the number of nodes required for training with 
\ourway. This leads to up to 50\% energy reduction. However, as seen in the figure, \flash~ has higher energy than the baseline system for \ifrmodelscaled~because of the low QPS (longer training time). Note that this high reduction is because we have a QPS target for training. If we were to compare absolute performance achieved by the baseline system and \ourway, the number of nodes running \ourway~ required would be higher, lowering power and energy reduction.
Figure \ref{fig:adsbaseline_energy} shows \adsmodel's average power consumption is reduced by up to 60\% in 1 node and 30\% in 2 node \ourway~ configurations. Nonetheless, because \ourway's QPS for  \adsmodel~ is considerably lower than the baseline config (longer training time), even when using 2 nodes running \ourway, energy consumption is higher compared to the baseline in all configurations, as seen in Figure \ref{fig:adsbaseline_energy}. 

\begin{figure}[]
 \centering
  \includegraphics[width=0.450\textwidth, trim={0.2cm 10.0cm 10.3cm 0.05cm },clip]{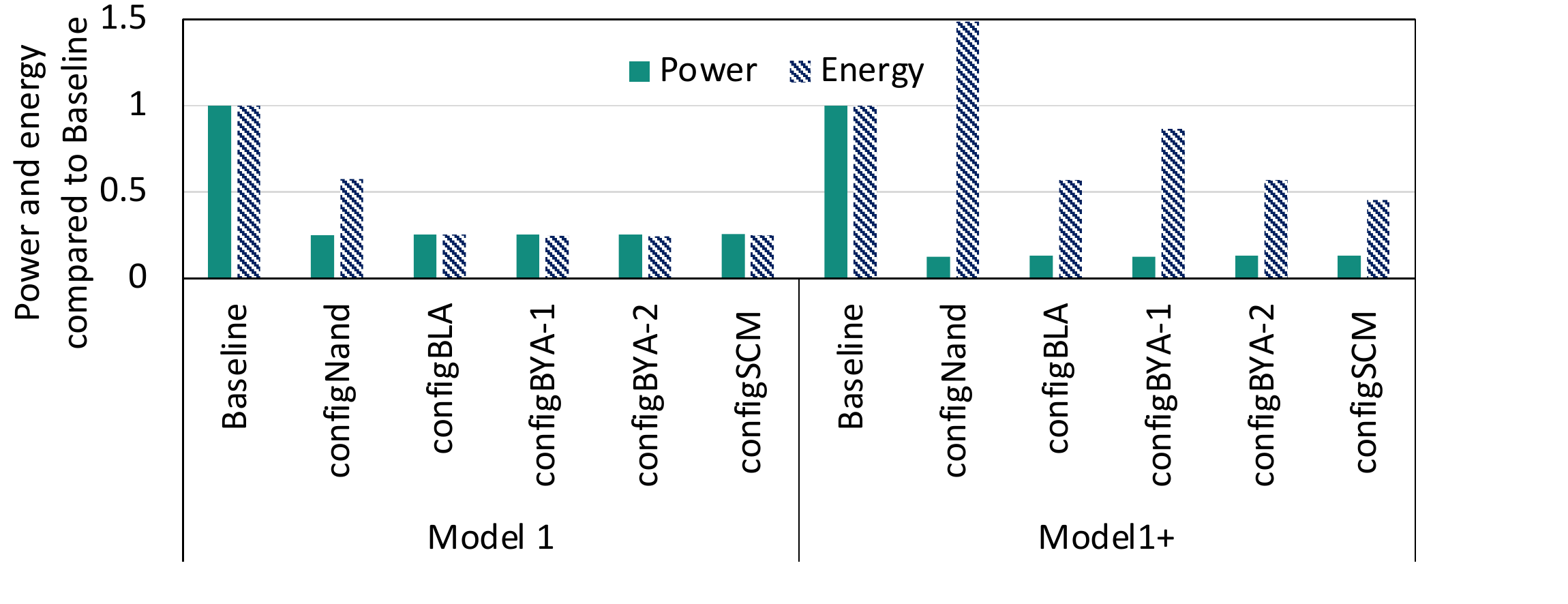}

 \caption{\footnotesize{Power and energy comparison of \ourway~ to baseline system running CLDRM+ for \ifrmodel~ and \ifrmodelscaled .}}
 \label{fig:baseline_energy}
\end{figure}
\begin{figure}[]
 \centering
  \includegraphics[width=0.50\textwidth, trim={0.2cm 9.5cm 11.3cm 0.05cm },clip]{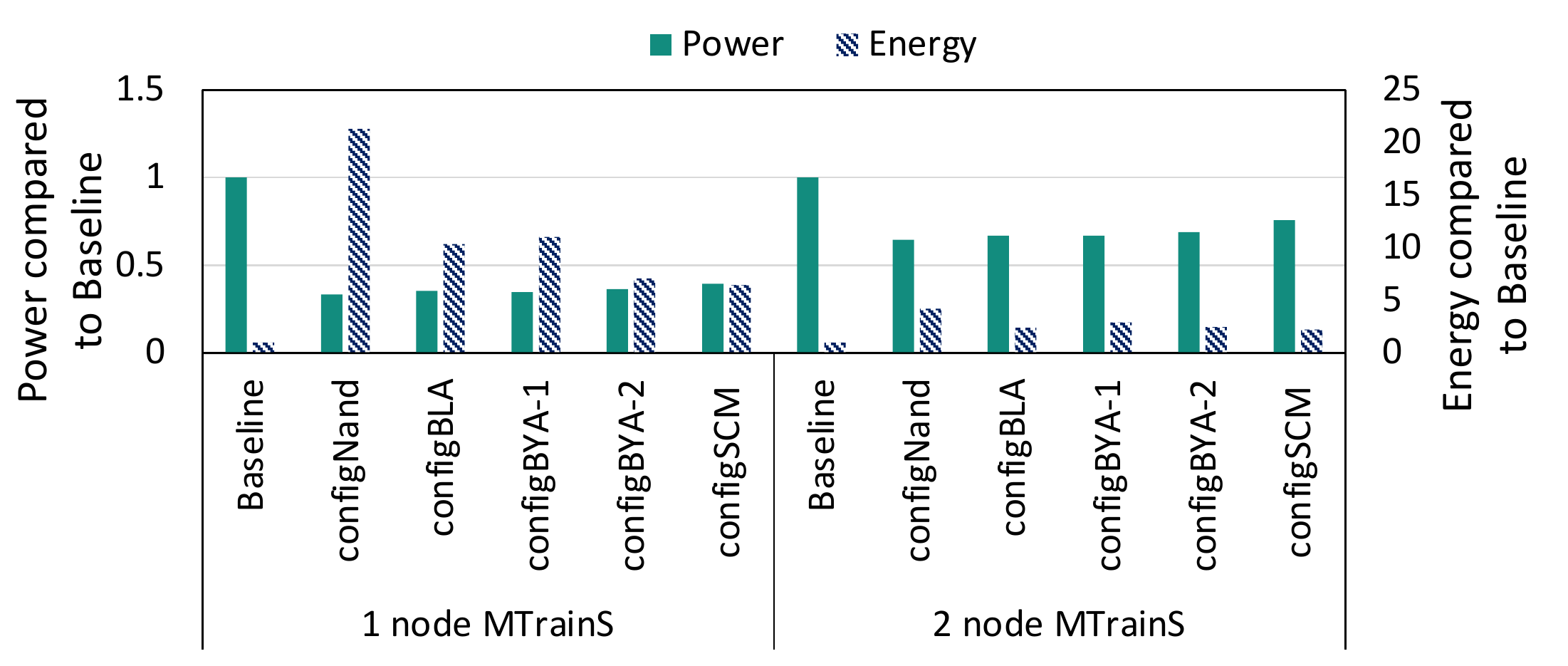}

 \caption{\footnotesize{Power and energy comparison of \ourway~ to baseline system running CLDRM+ for \adsmodel~ using 1 and 2 nodes eith MTrainS.}}
 \label{fig:adsbaseline_energy}
\end{figure}


\begin{figure}[]
 \centering
  \includegraphics[width=0.45\textwidth, trim={0.4cm 8.3cm 10.3cm 0.05cm },clip]{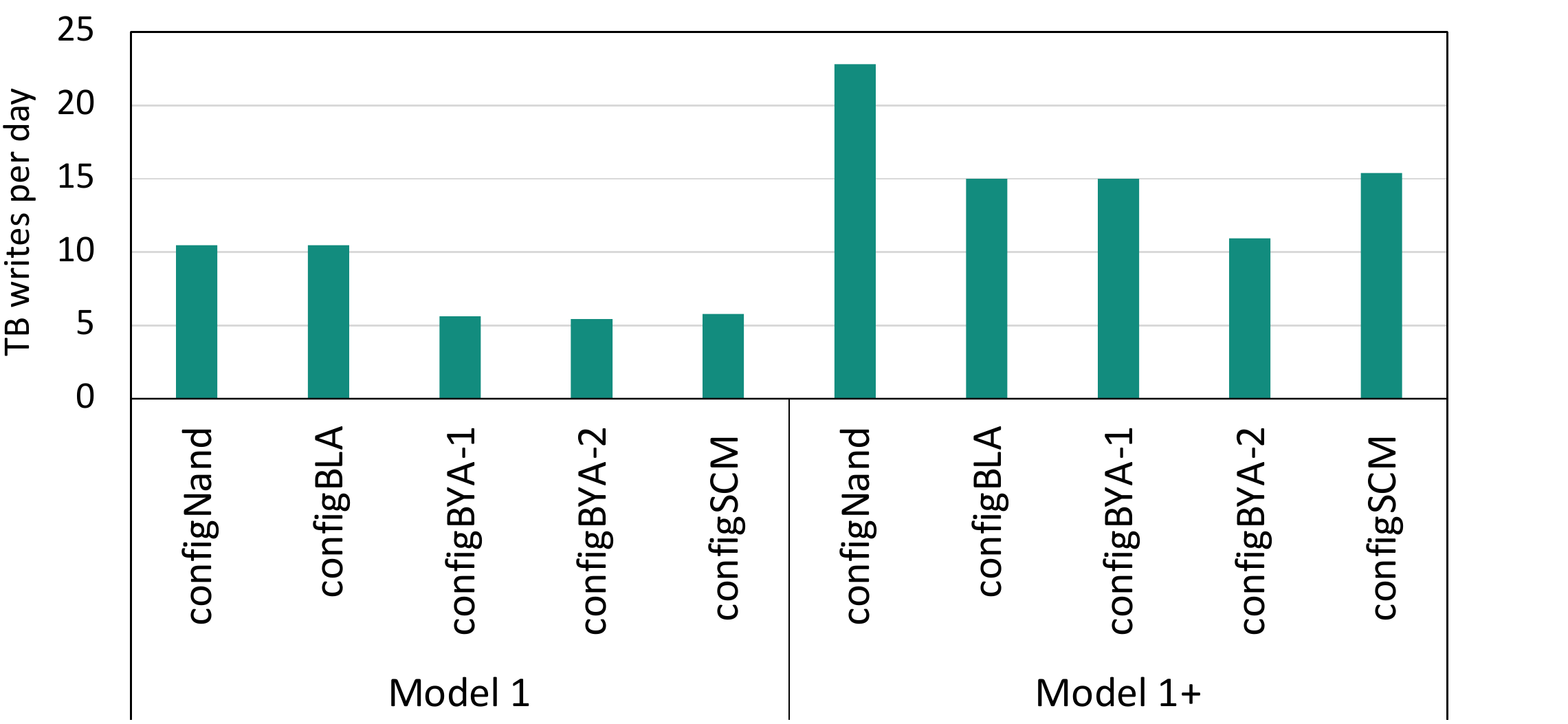}
 \caption{\footnotesize{Wear out comparison for \ifrmodel~ and \ifrmodelscaled .}}
 \label{fig:endurance}
\end{figure}




\subsection{Storage endurance and wear out}
In this and the following sections, we show detailed hardware characteristic analysis for the models that benefit from \ourway, i.e., \ifrmodel~ and \ifrmodelscaled. Figure \ref{fig:endurance} measures the TB written per day for \ifrmodel~ and \ifrmodelscaled~ to study if we meet our endurance requirements with the diverse configurations of \ourway. Based on the sizes of \optane~ and \flash, the endurances to avoid storage wear out are 200TB and 8TB data writes per day (DWPD), respectively. As seen in the figure, while we satisfy our QPS target with \sysOne~ for \ifrmodel, the write per day exceeds the endurance of \flash~(8TB). \sysThree~ and \sysFour~ satisfy the endurance because of the increase in cache size that decreases writes to storage. All configurations with \optane~ (\sysTwo ~ and \sysFive) meet the endurance because of its higher DWPD (200TB). For \ifrmodelscaled, all configurations backed by \flash~ do not meet the endurance. Therefore, for future models \optane~ is the best option to prevent storage wear out.

\subsection{Cache hit and IO utilization}
Figure \ref{figures:cache_hit} shows the cache hit rates for \ifrmodel~ and \ifrmodelscaled. For \ifrmodel~ in Figure \ref{figures:cache_hit_1x }, we observe 50\% cache hit rate with \flash~ and \sysTwo ~ that use only DRAM for cache, and addition of \pmem~ (\sysThree) increases the cache hit to 75\%. The 768GB \pmem~ configuration (\sysFour) does not increase the cache hit further because the 384GB \pmem~ is enough to capture the temporal locality. As a result, both configurations of \pmem~ reach similar QPS, as shown in Figure \ref{fig:qps_ifr}. For \ifrmodelscaled, as shown in Figure \ref{figures:cache_hit_2x}, we measure 40\% cache hit rate with \sysOne. The addition of \pmem~ increases the cache hit to 60\% and 70\% for \sysThree~ and \sysFour, respectively. In this case, as shown in Figure~\ref{fig:qps_ifr_scaled}, higher \pmem~ translates to higher QPS. The reason for the different behavior by \ifrmodel~and \ifrmodelscaled~is that the bigger embedding dimension in the latter model manifests higher pressure on the caches. As a result, a larger cache size can capture more locality, resulting in less IO traffic and improved QPS. 

As shown in Figure \ref{figure:iops}, the trainer achieves higher IOPS and bandwidth with \optane. In Figure \ref{figure:iops_1x} for \ifrmodel~ with \optane~ 1.5$\times$ more IOPS and 2.3$\times$ more effective BW ($IOPS \times embedding\_dim$) is measured, and consecutively more than 2$\times$ QPS compared to \flash, as shown in Figure \ref{fig:qps_ifr}. Adding \pmem~ decreases IOPS by up to 80\% and BW by up to 60\%. Here, the IO traffic is substantially reduced to the point that SSD traffic is not on the critical performance path. Figure \ref{figure:iops_2x} shows similar trends for \ifrmodelscaled~ with the main difference that the higher \pmem~ continues to improve QPS, and hence IOPS and effective BW. In summary, the \ourway~ configurations in Table \ref{tab:systems} improve QPS 1) by reducing IO traffic using \pmem~ and/or 2) by providing more IOPS using \optane.   

\begin{figure}[t!]

  \begin{subfigure}{.47\columnwidth}

\includegraphics[width=1.05\textwidth, trim={0.51cm 8.4cm 15.7cm 0.55cm },clip]{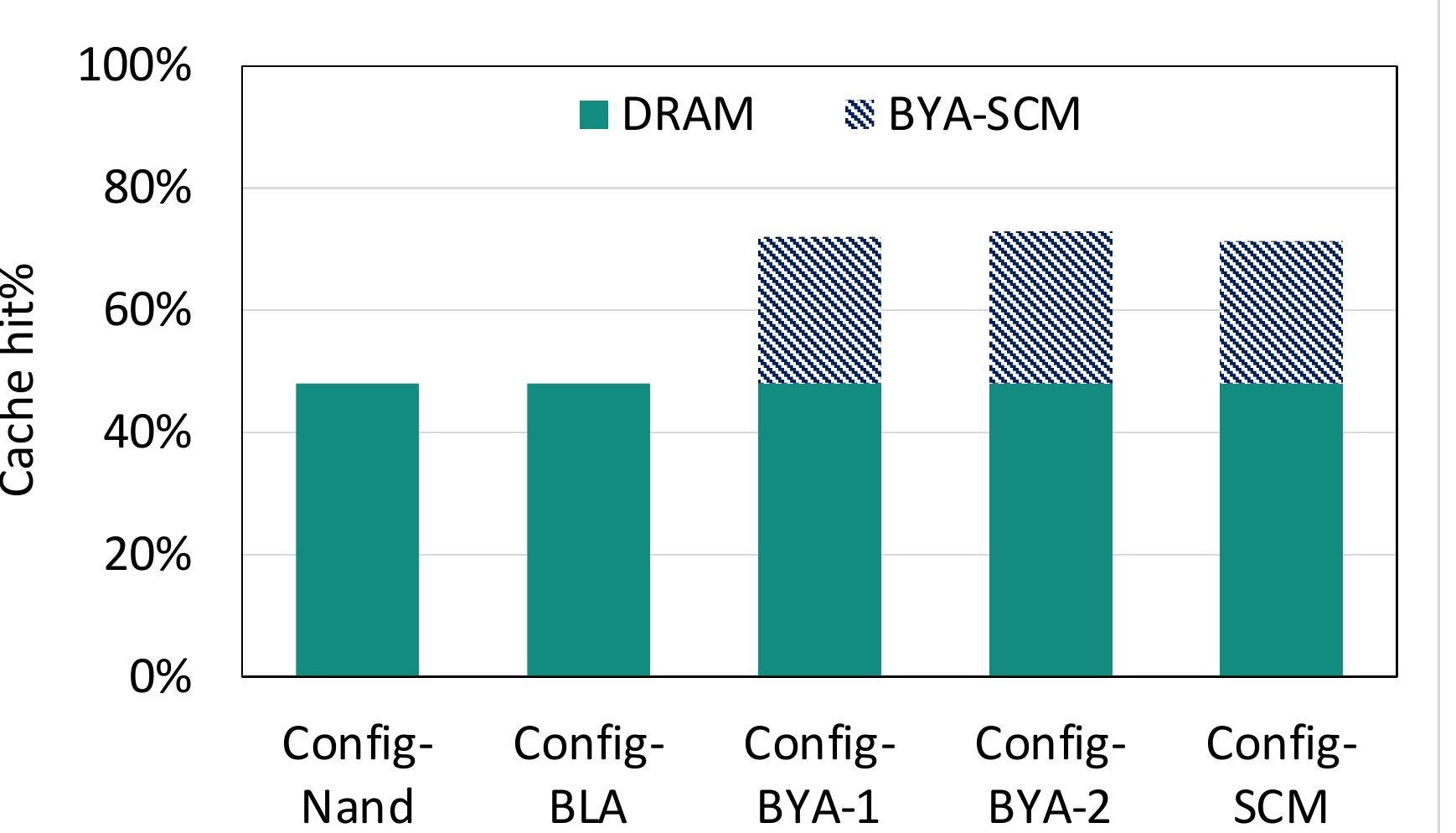}
\caption{\textmd{\ifrmodel} }
\label{figures:cache_hit_1x }
    \end{subfigure}
\begin{subfigure}{0.47\columnwidth}
\includegraphics[width=1.05\textwidth, trim={0.55cm 8.4cm 16.1cm 0.55cm },clip]{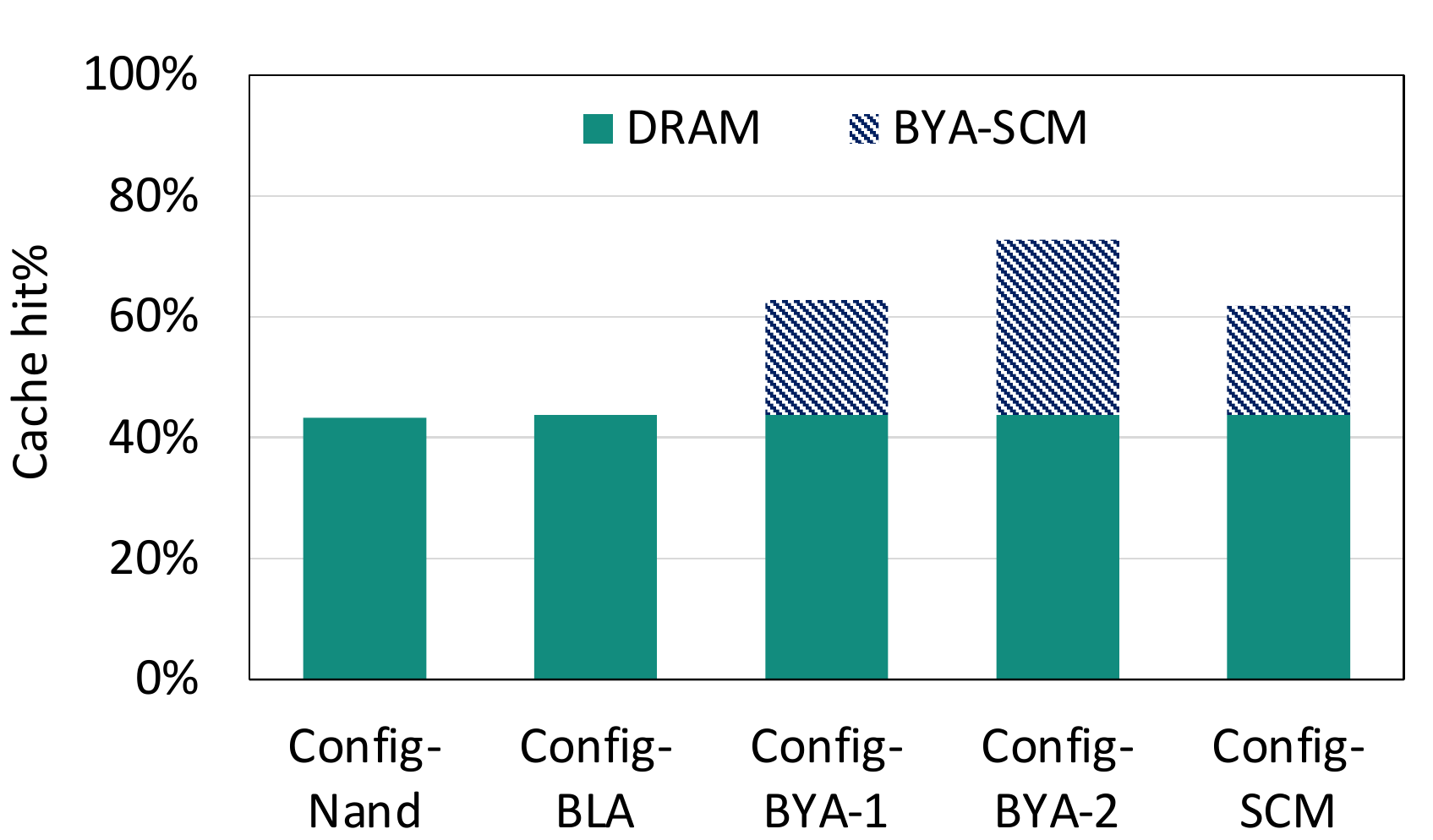}
\caption{\textmd{\ifrmodelscaled}}
\label{figures:cache_hit_2x}

    \end{subfigure}
\caption{
\footnotesize{Cache hit rates for \ifrmodel~ (a) and \ifrmodelscaled~(b) for different \ourway~ configurations.}}

\label{figures:cache_hit}
\end{figure}

\begin{figure}[t!]

  \begin{subfigure}{.475\columnwidth}
\includegraphics[width=1.15\textwidth, trim={0.150cm 9.1cm 15.0cm 0.005cm },clip]{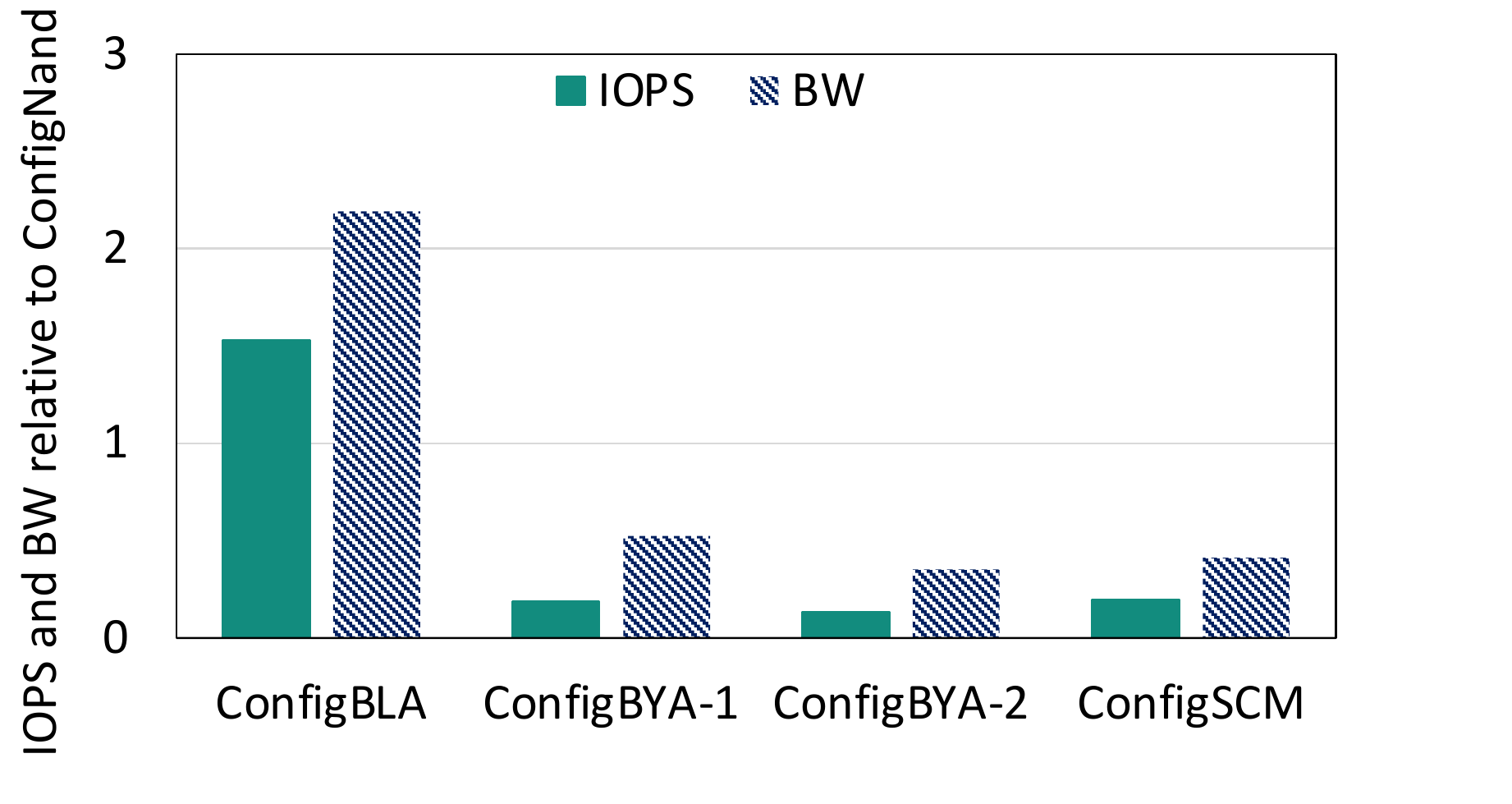}

\caption{\textmd{\ifrmodel} }
    \label{figure:iops_1x}
    \end{subfigure}
\begin{subfigure}{0.475\columnwidth}
 \includegraphics[width=1.1\textwidth, trim={0.15cm 9.1cm 16.0cm 0.005cm },clip]{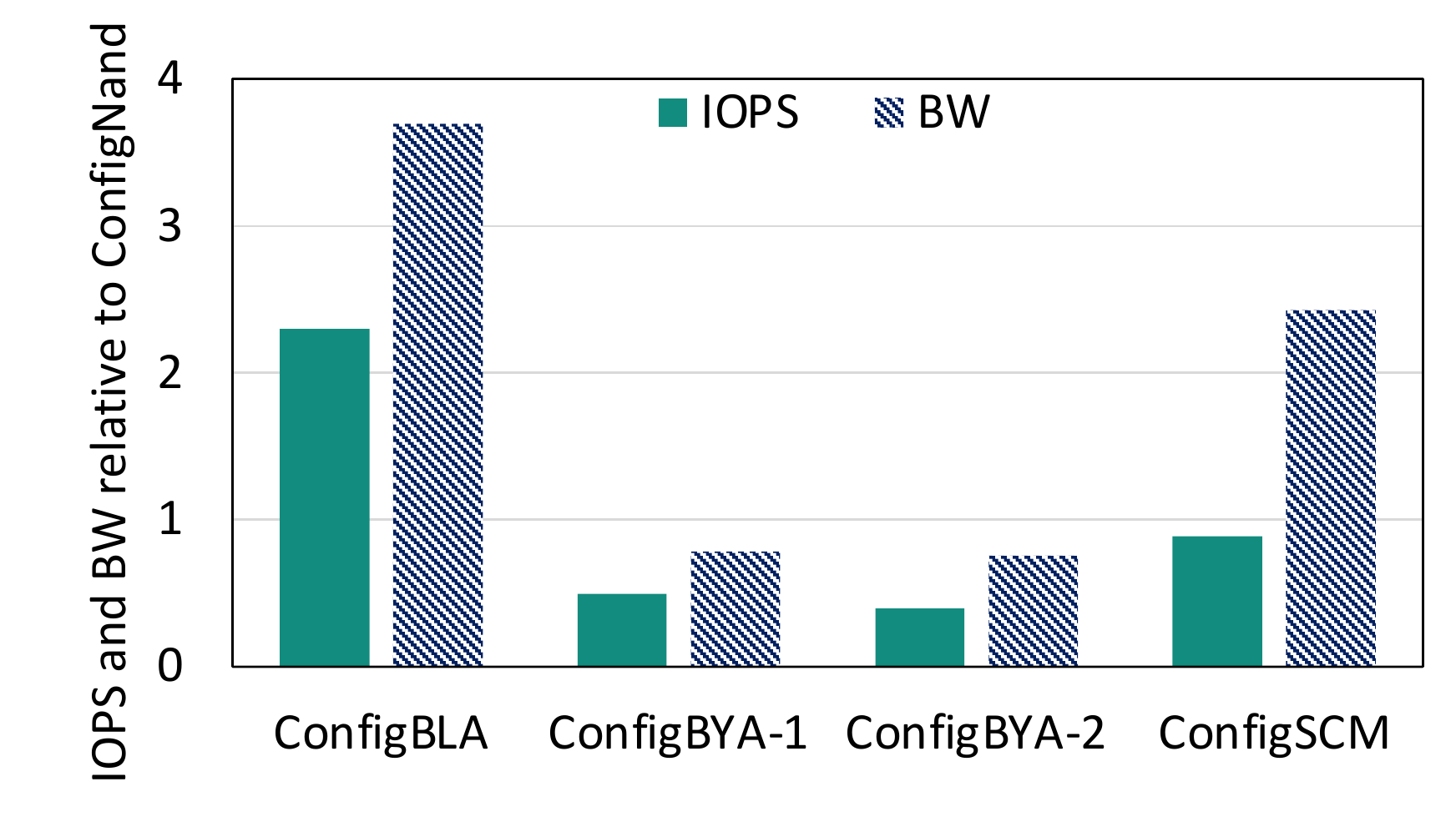}

\caption{\textmd{\ifrmodelscaled}}

    \label{figure:iops_2x}
    \end{subfigure}
\caption{\footnotesize{
IOPS and effective SSD BW for \ifrmodel~and \ifrmodelscaled. }
}
    \label{figure:iops}
\end{figure}



\vspace{-2mm}
\subsection{Embedding table assignment efficiency}\label{pl}
In this section, we compare the benefit of four embedding table placements for \ifrmodel~and \ifrmodelscaled~ running in our config with the largest resource, \sysFour. The first is placing all embedding tables in RocksDB storage (\flash~ and \optane) and using the available size of DRAM and \pmem~ memories for caching. We use this as the baseline in Figure \ref{fig:placement} and call it unoptimized because table placement is only optimized for maximizing the utilization of the sizes of SSDs instead of BW; hence, the data access volume is not balanced among the GPUs. When we compare the unoptimized with using BW to balance the data access of the GPUs, we increase QPS by 15\%. Note that using just SSD with DRAM and \pmem~ caching has lower HBM utilization. We then compare this to applying a linear programming solver with only the size of embedding tables and memory types as input. We get 2.5-3.5$\times$ more QPS with this placement. Using both size and bandwidth aware placement further increases the QPS to 3.2-4.2$\times$. Hence, table placement is critical when training DLRM with heterogeneous memories, and considering both the size and BW provides the best performance.



\section{Related work}\label{related_work}
\textbf{Memory capacity extension for recommendation systems:}
The high memory capacity and BW demand of embedding operation in recommendation systems impose a challenge on the memory system. Training and Inference use cases present their own unique challenges. For example, in Inference, the latency of each query is important. Ardestani \etal~~\cite{ardestani2021supporting} present an end-to-end system to leverage SSDs while keeping the latency manageable, and Wilkening \etal~~\cite{dlrm_ssd} use the controller in SSD to offload some of the compute closer to the data. Eisenman \etal~ ~\cite{eisenman2019bandana} uses SCM to increase memory capacity per host. However, it requires offline preprocessing of embedding tables, which is not applicable for training.

Training is less sensitive to latency but requires higher BW and frequent parameter updates, and hence read and write traffic to the SSDs. Zhao \etal~~\cite{zhao2020distributed} present a training system that leverages HBM, DRAM, and SSD. They leverage pipelining to hide SSD latency and use caching to hide lower SSD BW. They follow a parameter server scheme for training large models, as opposed to our distributed, synchronized scheme. Also, in contrast, to sustain high throughput, we leverage the GPU for cache management. Additionally, we show different SCM technologies' performance and power impact on training. Balasubramanian \etal~ ~\cite{balasubramanian2021cdlrm} implements a CPU-managed cache to leverage HBM and DRAM to expand the memory capacity of embedding tables but only extends to DRAM. Yang \etal~~\cite{yang2020mixedprecision} propose and implement a software caching scheme in GPU backed by CPU memory for embeddings. We build on a similar scheme as ~\cite{yang2020mixedprecision} and enable multi-level caching with DRAM and \pmem, backed by SSDs. However, while they also use caching with DLRM, they focus on using different precisions to reduce embedding storage sizes. In contrast, we extend the memory hierarchy to \pmem~ and flash to expand memory per host and remain neutral to the accuracy of training.   

\begin{figure}[t!]
 \centering
  \includegraphics[width=0.34\textwidth, trim={0.5cm 9.0cm 8.8cm 0.05cm },clip]{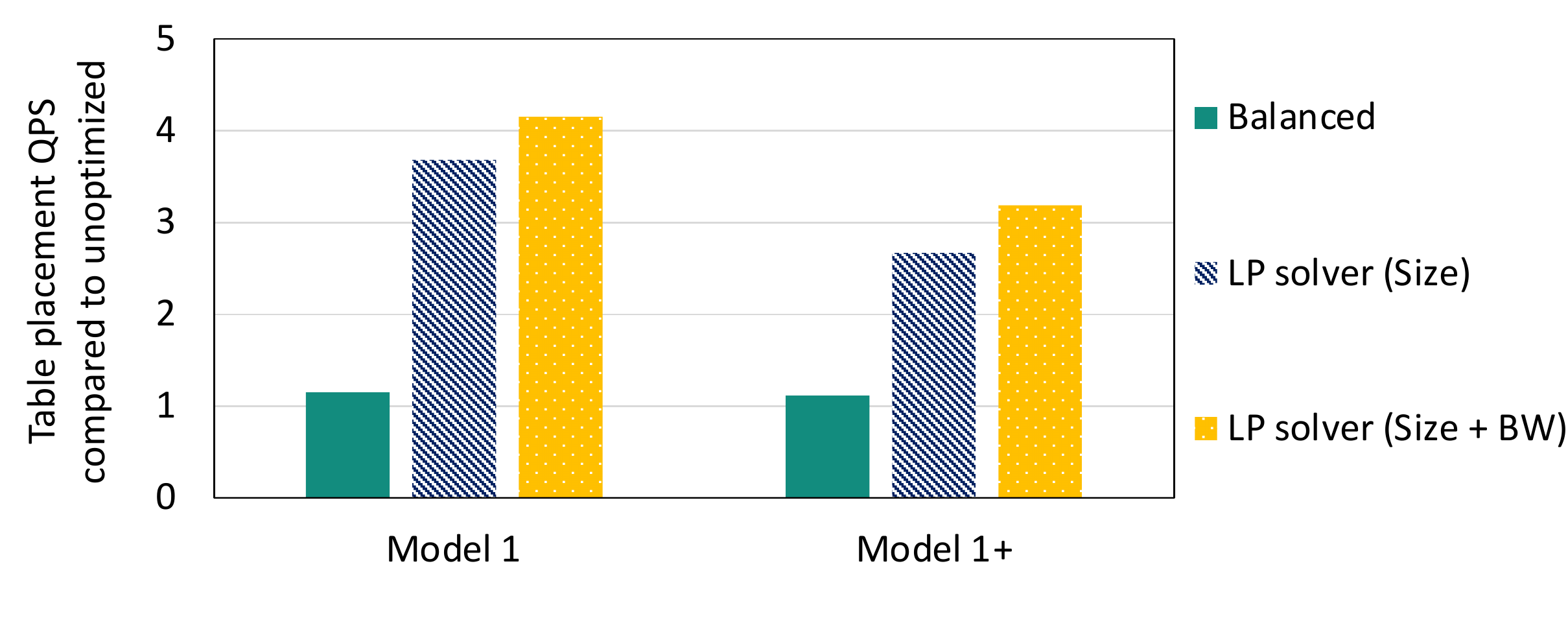}

 \caption{\footnotesize{QPS increase for different types of embedding table placement for \ifrmodel~and \ifrmodelscaled}}
 \label{fig:placement}
\end{figure}


\noindent\textbf{SCM usage for AI workloads:} 
Hildebrand \etal~\cite{pmem_AI} developed an integer linear programming-based system that moves tensors between DRAM and SCM. They show that optimized data placement achieves higher performance than a naive approach. Similarly, in our designs, based on the characteristics of applications, we carefully place data on different memories. 

\section{Conclusion}\label{conclusion}

In the pursuit of better model quality, recommendation model complexity, size, and training data are increasing, which imposes considerable pressure on the underlying platforms' compute, bandwidth, IO, and memory capacity. This paper tackles the pressure on the memory capacity and presents a hierarchical memory-based trainer \ourway, to increase memory capacity per host. We quantitatively compare the performance and system-level implications of byte and block addressable SCMs. Based on the temporal locality in embedding tables across large-scale production workloads, we apply a multi-level cache, utilizing SCM to reduce the number of nodes required to load and train the memory-bound models by 1/4-1/8th. Adding SCM to a platform results in a minimal increase in platform cost and power, given that other components, such as GPU, dominate the power. As a result, we observe overall efficiency benefits due to the fewer nodes required to train a model. 
There are limits to this approach for bandwidth-bound models. In these cases, the multi-level cache fails to deliver enough bandwidth to satisfy the model performance requirements under the hardware configurations evaluated. 




\bibliographystyle{plain}

\bibliography{reference.bib}

\end{document}
